\documentclass[12pt,preprint]{aastex}
\usepackage{graphicx,amsmath,amssymb}
\usepackage{pdflscape}
\usepackage{url}

\newcommand{\bdv}[1]{\mbox{\boldmath$#1$}}

\def\au{{\rm AU}}

\def\muas{\mu{\rm as}}

\def\rel{{\rm rel}}

\def\e{{\rm E}}
\def\bpi{{\bdv\pi}}

\begin{document}
\title{First simultaneous microlensing observations by two space telescopes:
{\it Spitzer} \& {\it Swift} reveal a brown dwarf in event OGLE-2015-BLG-1319}
 
\author{
Y. Shvartzvald$^{1,a}$,
Z. Li$^{2}$,
A. Udalski$^{3}$,
A. Gould$^{4}$,
T.~Sumi$^{5}$,
R.~A.~Street$^{2}$,
S.~Calchi~Novati$^{6,7}$,
M. Hundertmark$^{8}$,
V. Bozza$^{9,10}$,\\
and\\
C. Beichman$^{11}$,
G. Bryden$^{1}$,
S. Carey$^{12}$,
J. Drummond$^{13}$,
M. Fausnaugh\altaffilmark{4},
B.~S.~Gaudi\altaffilmark{4},
C.~B.~Henderson$^{1,a}$,
T.G. Tan$^{14}$,
B. Wibking\altaffilmark{4},
R. W. Pogge$^{4}$,
J.~C.~Yee$^{15,b}$,
W. Zhu$^{4}$,\\
($Spitzer$ team)\\
and\\
Y. Tsapras$^{16}$,
E. Bachelet$^{2,17}$,
M. Dominik$^{18,c}$,
D.~M.~Bramich$^{17}$,
A. Cassan$^{19}$,
R.~Figuera~Jaimes$^{18,20}$,
K. Horne$^{18}$,
C. Ranc$^{19}$,
R.~Schmidt$^{16}$,
C. Snodgrass$^{21}$,
J.~Wambsganss$^{16}$,
I. A. Steele$^{22}$,
J. Menzies$^{23}$,
S.~Mao$^{24,25,26}$,\\
(RoboNet)\\
and\\
R. Poleski$^{4,3}$,
M. Pawlak$^{3}$,
M.\,K. Szyma{\'n}ski$^{3}$,
J. Skowron$^{3}$
P. Mr{\'o}z$^{3}$,
S.~Koz{\l}owski$^{3}$,
{\L}.~Wyrzykowski$^{3}$,
P. Pietrukowicz$^{3}$,
I.~Soszy{\'n}ski$^{3}$,
K. Ulaczyk$^{27}$,\\
(OGLE group)\\
and\\
F.~Abe$^{28}$,
Y. Asakura$^{28}$,
R.~K.~Barry$^{29}$,
D.~P.~Bennett$^{30,31}$,
A.~Bhattacharya$^{30}$,
I.A.~Bond$^{32}$,
M.~Freeman$^{33}$,
Y. Hirao$^{5}$,
Y.~Itow$^{28}$,
N.~Koshimoto$^{5}$,
M.C.A. Li$^{33}$,
C.H.~Ling$^{32}$,
K.~Masuda$^{28}$,
A.~Fukui$^{34}$,
Y.~Matsubara$^{28}$,
Y.~Muraki$^{28}$,
M. Nagakane$^{5}$,
T. Nishioka$^{35}$,
K.~Ohnishi$^{36}$,
H. Oyokawa$^{28}$,
N.~J.~Rattenbury$^{33}$,
To.~Saito$^{37}$,
A. Sharan$^{33}$,
D.J.~Sullivan$^{38}$,
D.~Suzuki$^{30,31}$,
P.~J.~Tristram$^{39}$,
A.~Yonehara$^{35}$,\\
(MOA group)\\
and\\
U. G. J{\o}rgensen$^{8}$,
M. J. Burgdorf$^{40}$,
S.~Ciceri$^{41}$,
G. D'Ago$^{9,10,42}$,
D. F. Evans$^{43}$, 
T.~C.~Hinse$^{44}$, 
N. Kains$^{26}$,  
E.~Kerins$^{26}$, 
H. Korhonen$^{45,8,46}$,
L. Mancini$^{41}$,
A. Popovas$^{8}$, 
M.~Rabus$^{47}$,
S.~Rahvar$^{48}$,
G. Scarpetta$^{9,42}$,
J. Skottfelt$^{49,8}$,
J. Southworth$^{43}$,
N.~Peixinho$^{50,51}$,
P. Verma$^{42}$,\\
(MiNDSTEp)\\
and\\
B. Sbarufatti$^{52,53}$,
J.~A.~Kennea$^{53}$,
N. Gehrels$^{54}$\\
($Swift$)\\
}
\altaffiltext{1}{Jet Propulsion Laboratory, California Institute of Technology, 4800 Oak Grove Drive, Pasadena, CA 91109, USA}
\altaffiltext{2}{Las Cumbres Observatory Global Telescope Network, 6740 Cortona Drive, suite 102, Goleta, CA 93117, USA}
\altaffiltext{3}{Warsaw University Observatory, Al.~Ujazdowskie~4, 00-478~Warszawa, Poland}
\altaffiltext{4}{Department of Astronomy, Ohio State University, 140 W. 18th Ave., Columbus, OH  43210, USA}
\altaffiltext{5}{Department of Earth and Space Science, Osaka University, Osaka 560-0043, Japan}
\altaffiltext{6}{Infrared Processing and Analysis Center and NASA Exoplanet Science Institute, California Institute of Technology, Pasadena, CA 91125, USA}
\altaffiltext{7}{Dipartimento di Fisica ``E. R. Caianiello'', Universit\`a di Salerno, Via Giovanni Paolo II, 84084 Fisciano (SA),\ Italy}
\altaffiltext{8}{Niels Bohr Institute \& Centre for Star and Planet Formation, University of Copenhagen, {\O}ster Voldgade 5, 1350 Copenhagen, Denmark}
\altaffiltext{9}{Dipartimento di Fisica "E.R. Caianiello", Universit{\`a} di Salerno, Via Giovanni Paolo II 132, 84084, Fisciano, Italy}
\altaffiltext{10}{Istituto Nazionale di Fisica Nucleare, Sezione di Napoli, Napoli, Italy}
\altaffiltext{11}{NASA Exoplanet Science Institute, California Institute of Technology, Pasadena, CA 91125, USA}
\altaffiltext{12}{$Spitzer$, Science Center, MS 220-6, California Institute of Technology,Pasadena, CA, USA}
\altaffiltext{13}{Possum Observatory, Patutahi, NZ}
\altaffiltext{14}{Perth Exoplanet Survey Telescope, Perth, Australia}
\altaffiltext{15}{Harvard-Smithsonian Center for Astrophysics, 60 Garden St., Cambridge, MA 02138, USA}
\altaffiltext{16}{Astronomisches Rechen-Institut, Zentrum f{\"u}r Astronomie der Universit{\"a}t Heidelberg (ZAH), 69120 Heidelberg, Germany}
\altaffiltext{17}{Qatar Environment and Energy Research Institute(QEERI), HBKU, Qatar Foundation, Doha, Qatar}
\altaffiltext{18}{SUPA, School of Physics \& Astronomy, University of St Andrews, North Haugh, St Andrews KY16 9SS, UK}
\altaffiltext{19}{Sorbonne Universit\'es, UPMC Univ Paris 6 et CNRS, UMR 7095, Institut d'Astrophysique de Paris, 98 bis bd Arago, 75014 Paris, France}
\altaffiltext{20}{European Southern Observatory, Karl-Schwarzschild-Str. 2, 85748 Garching bei M\"unchen, Germany}
\altaffiltext{21}{Planetary and Space Sciences, Department of Physical Sciences, The Open University, Milton Keynes, MK7 6AA, UK}
\altaffiltext{22}{Astrophysics Research Institute, Liverpool John Moores University, Liverpool CH41 1LD, UK}
\altaffiltext{23}{South African Astronomical Observatory, PO Box 9, Observatory 7935, South Africa}
\altaffiltext{24}{Physics Department and Tsinghua Centre for Astrophysics, Tsinghua University, Beijing 100084, China}
\altaffiltext{25}{National Astronomical Observatories, Chinese Academy of Sciences, 20A Datun Road, Chaoyang District, Beijing 100012, China}
\altaffiltext{26}{Jodrell Bank Centre for Astrophysics, School of Physics and Astronomy, The University of Manchester, Oxford Road, Manchester M13 9PL, UK}
\altaffiltext{27}{Department of Physics, University of Warwick, Gibbet Hill Road, Coventry, CV4 7AL, UK}
\altaffiltext{28}{Institute for Space-Earth Environmental Research, Nagoya University, Nagoya 464-8601, Japan}
\altaffiltext{29}{Astrophysics Science Division, NASA Goddard Space Flight Center, Greenbelt, MD 20771, USA}
\altaffiltext{30}{University of Notre Dame, Department of Physics, 225 Nieuwland Science Hall, Notre Dame, IN 46556-5670, USA}
\altaffiltext{31}{Laboratory for Exoplanets and Stellar Astrophysics, NASA/Goddard Space Flight Center, Greenbelt, MD 20771, USA}
\altaffiltext{32}{Institute of Information and Mathematical Sciences, Massey University, Private Bag 102-904, North Shore Mail Centre, Auckland, New Zealand}
\altaffiltext{33}{Department of Physics, University of Auckland, Private Bag 92-019, Auckland 1001, New Zealand}
\altaffiltext{34}{Okayama Astrophysical Observatory, National Astronomical Observatory of Japan, Asakuchi, Okayama 719-0232, Japan}
\altaffiltext{35}{Department of Physics, Faculty of Science, Kyoto Sangyo University, 603-8555 Kyoto, Japan}
\altaffiltext{36}{Nagano National College of Technology, Nagano 381-8550, Japan}
\altaffiltext{37}{Tokyo Metropolitan College of Aeronautics, Tokyo 116-8523, Japan}
\altaffiltext{38}{School of Chemical and Physical Sciences, Victoria University, Wellington, New Zealand}
\altaffiltext{39}{Mt. John University Observatory, P.O. Box 56, Lake Tekapo 8770, New Zealand}
\altaffiltext{40}{Meteorologisches Institut, Universit{\"a}t Hamburg, Bundesstra\ss{}e 55, 20146 Hamburg, Germany}
\altaffiltext{41}{Max Planck Institute for Astronomy, K{\"o}nigstuhl 17, 69117 Heidelberg, Germany}
\altaffiltext{42}{Istituto Internazionale per gli Alti Studi Scientifici (IIASS), Via G. Pellegrino 19, 84019 Vietri sul Mare (SA), Italy}
\altaffiltext{43}{Astrophysics Group, Keele University, Staffordshire, ST5 5BG, UK}
\altaffiltext{44}{Korea Astronomy \& Space Science Institute, 776 Daedukdae-ro, Yuseong-gu, 305-348 Daejeon, Republic of Korea}
\altaffiltext{45}{Finnish Centre for Astronomy with ESO (FINCA), V{\"a}is{\"a}l{\"a}ntie 20, FI-21500 Piikki{\"o}, Finland}
\altaffiltext{46}{Niels Bohr Institute \& Dark Cosmology Centre, University of Copenhagen, Juliane Mariesvej 30, 2100 - Copenhagen {\O}, Denmark}
\altaffiltext{47}{Instituto de Astrof\'isica, Facultad de F\'isica, Pontificia Universidad Cat\'olica de Chile, Av. Vicu\~na Mackenna 4860, 7820436 Macul, Santiago, Chile}
\altaffiltext{48}{Department of Physics, Sharif University of Technology, PO Box 11155-9161 Tehran, Iran}
\altaffiltext{49}{Centre for Electronic Imaging, Department of Physical Sciences, The Open University, Milton Keynes, MK7 6AA, UK}
\altaffiltext{50}{CITEUC -- Centre for Earth and Space Science Research of the University of Coimbra, Observat\'orio Geof\'isico e Astron\'omico da U.C., 3030-004 Coimbra, Portugal}
\altaffiltext{51}{Unidad de Astronom\'ia, Fac. de Cs. B\'asicas, Universidad de Antofagasta, Avda U. de Antofagasta 02800, Antofagasta, Chile}
\altaffiltext{52}{Istituto Nazionale di Astrofisica, Osservatorio Astronomico di Brera, Via E. Bianchi 46, I-23807, Merate, Italy}
\altaffiltext{53}{Department of Astronomy and Astrophysics, Pennsylvania State University, University Park, PA, 16802, USA}
\altaffiltext{54}{Astrophysics Science Division, NASA Goddard Space Flight Center, Greenbelt, MD, 20740, USA}
\altaffiltext{a}{NASA Postdoctoral Program Fellow}
\altaffiltext{b}{Sagan Fellow}
\altaffiltext{c}{Royal Society University Research Fellow}

\begin{abstract}
Simultaneous observations of microlensing events from multiple locations allow for the 
breaking of degeneracies between the physical properties of the lensing system, specifically by exploring different regions of the lens plane
and by directly measuring the ``microlens parallax''.
We report the discovery of a 30--55$M_J$ brown dwarf orbiting a K dwarf 
in microlensing event OGLE-2015-BLG-1319.
The system is located at a distance of $\sim$5 kpc toward the Galactic bulge.
The event was observed by several ground-based groups as well as by $Spitzer$ and $Swift$, allowing the measurement of the physical properties.
However, the event is still subject to an 8-fold degeneracy, in particular the well-known close-wide degeneracy,
and thus the projected separation between the two lens components is either $\sim$0.25 AU or $\sim$45 AU.
This is the first microlensing event observed by $Swift$, with the UVOT camera.
We study the region of microlensing parameter space to which $Swift$ is sensitive, finding that while for this event $Swift$ could not measure the microlens parallax
with respect to ground-based observations, it can be important for other events. Specifically, for detecting nearby brown dwarfs
and free-floating planets in high magnification events.

\end{abstract}

\keywords{gravitational lensing: micro -- binaries: general -- stars: brown dwarfs --
Galaxy: bulge}

\section{{Introduction}
\label{sec:intro}}

The $Spitzer$ 2014 and 2015 microlensing campaigns have revolutionized the microlensing field.
The satellite observations of over 200 events that were discovered and monitored simultaneously by ground-based surveys
facilitated the systematic measurement of the microlens parallax, $\pi_\e$, for the majority of the events -- a crucial quantity for determining the physical properties of the lensing system.
Simultaneous ground and space observations of microlensing events were conducted only twice prior to these campaigns,
once with $Spitzer$ \citep{Dong.2007.A}, and once with the {\it Deep Impact} (or $EPOXI$) spacecraft \citep{Muraki.2011.A}.
The 2014-2015 $Spitzer$ campaigns have already led to the detection of two planets (\citealt{Udalski.2015.A,Street.2016.A}),
the first caustic-crossing binary-lens event with a satellite parallax measurement \citep{Zhu.2015.B},
a massive remnant in a wide binary (\citealt{Shvartzvald.2015.A}), and mass measurements of isolated objects
(\citealt{Zhu.2015arXiv.A}) -- one of which is a brown dwarf (BD).
In addition, $Spitzer$ observations allowed \cite{Bozza.2016.A} to break a strong planet/binary degeneracy in the event OGLE-2015-BLG-1212,
finding that the companion is a low-mass star and not a planet.
These campaigns are the first steps for measuring the Galactic distribution of planets (\citealt{Calchi.2015.A}),
a demographic regime that can currently only be explored by microlensing.

A microlensing event is characterized by the Einstein timescale
$t_\e$, which combines three physical properties of the lens-source
system,
\begin{equation}
t_\e = {\theta_\e\over\mu};
\qquad
\theta_\e^2\equiv \kappa M \pi_\rel;
\qquad
\kappa\equiv {4 G\over c^2\au}\simeq 8.14{{\rm mas}\over M_\odot}.
\label{eqn:tedef}
\end{equation}
Here $\theta_\e$ is the angular Einstein radius, $M$ is the total lens mass,
$\pi_\rel = \au(D_L^{-1}-D_S^{-1})$ is the lens-source relative parallax, and
$\mu$ is the lens-source relative proper motion.
Although the timescale will be approximately the same, the light curve of a microlensing event as seen from two (or more) separated observers (e.g., from Earth and space) is different,
due to either a different observed source trajectory, a
time shift, or both \citep{Refsdal.1966.A,Gould.1994.A}.
Since the physical separation between the two observers ($\bf D_\perp$) is known, this directly yields the microlens parallax,
\begin{equation}
\bpi_\e = \frac{\au}{D_\perp}(\Delta\tau,\Delta\beta);
\qquad \Delta\tau = \frac{t_{0,\rm sat} - t_{0,\oplus} }{t_\e};
\qquad \Delta\beta = \pm u_{0,\rm sat} - \pm u_{0,\oplus},
\label{eqn:pieframe}
\end{equation}
where the subscripts indicate parameters as measured from the satellite
and Earth.  Here, $(t_0,u_0,t_\e)$ are the standard ``Paczy{\'n}ski'' point-lens
microlensing parameters: time of minimal separation between the source and the lens, 
the impact parameter in angular Einstein radius units, and the event timescale.
Different observed source trajectories
will increase the probability of detecting companions to the lens star,
by exploring different regions of the lens plane (\citealt{Zhu.2015.A}).
However, due to the symmetry of the problem it usually suffers from a four-fold degeneracy (in $\Delta\beta$, see Eq.~\ref{eqn:pieframe}).
As pointed out by \cite{Refsdal.1966.A} and \cite{Gould.1994.A}, observing the same event from a third location can resolve this problem.
If the third observer is not on the 
projected line (with respect to the lensing event) defined by the first two observers,
it can completely remove the microlens parallax degeneracy, both in magnitude and direction. Even if the third observer is on the same projected line but has a different separation than the first two,
it will likely
view a different source trajectory.
In addition, it will be sensitive to different microlens parallax magnitudes (if the separation is too large, the magnification for one of the 
observers can be too low for the event to be detected).

Here we present the analysis of OGLE-2015-BLG-1319.
This is the first microlensing event observed by two space telescopes, $Spitzer$ and $Swift$, and from ground.
The ground light-curve shows a short anomaly over the peak due to a companion, and the parallax measurement from $Spitzer$ allow us to determine that it is a BD.
This additional BD detection provides supporting evidence for a conclusion previously drawn from microlensing studies,
that BDs might be common at separations of a few AU (\citealt{Shvartzvald.2016.A}). 
The small separation of $Swift$ from Earth did not allow for an independent measurement of the microlens parallax for this specific event.
However, as we discuss further below, it might be possible to measure the microlens parallax with $Swift$ alone in other events.

The paper is arranged as follows: we describe the observations from the ground-based observatories and from $Spitzer$ and $Swift$ in Section~\ref{sec:obs}.
In Section~\ref{sec:anal}, we present the microlensing model and try to resolve the degeneracy of the projected separation between the companion and its host. 
In Section~\ref{sec:physical}, we use the color-magnitude diagram
(CMD) to characterize the source properties and combine them with the microlensing model to derive the lens physical properties.
We study the feasibility of using $Swift$ to measure the microlens parallax in Section~\ref{sec:swift}. Finally, in Section~\ref{sec:discussion},
we summarize our results.

{\section{Observational data and reduction}
\label{sec:obs}}

\subsection{Ground observations}

The microlensing event OGLE-2015-BLG-1319 was first alerted on
June 11, 2015, 19:44 UT by the Optical Gravitational Lens Experiment (OGLE),
which operates the 1.3m Warsaw telescope at the Las Campanas Observatory in Chile (\citealt{Udalski.2015.B}),
using the OGLE Early Warning System (EWS, \citealt{Udalski.2003.A}).
At equatorial coordinates RA = 17:57:46.4, Dec = $-$32:28:19.9 (J2000.0),
the event lies in OGLE field BLG508,
which has a relatively low observing cadence of 0.5--1 times per night.
Most observations were in $I$ band, with additional sparse $V$ band observations for
source characterization.
OGLE photometry was extracted by their standard difference image analysis (DIA) procedure (\citealt{Udalski.2003.A}).

The event was also observed by the Microlensing Observations in Astrophysics (MOA) collaboration,
who operate the 1.8m MOA-II telescope at the Mt. John Observatory in New Zealand (\citealt{Sumi.2003.A}),
and designated as MOA-2015-BLG-292. Observations were in the ``MOA-Red'' filter (a wide $R/I$ filter), with a cadence of
1--5 times per night. The MOA data were reduced using their routine DIA procedure (\citealt{Bond.2001.A}).

The $Spitzer$ team selected and announced OGLE-2015-BLG-1319 as a $Spitzer$ target
on June 25 UT 2:00 (HJD'=7198.58) and aggressively alerted it as being an extreme high-magnification event
in the following days.
Based on these alerts, sustained follow-up observations were carried out.
First, the RoboNet team observed the event using 5 telescopes from the Las Cumbres Observatory Global Telescope (LCOGT) in Chile, South
Africa, and Australia. These observations were designed to increase the planet sensitivity by obtaining continuous coverage of the entire peak region.
Most observations were in $I$ band, and a few with $V$ band.
While these were not used for the source color characterization, they allow for a better coverage of the light curve.
LCOGT data were reduced using DanDIA \citep{Bramich.2008.A}.
The MiNDSTEp team followed the event using the Danish 1.54-m telescope hosted at ESO's La Silla observatory in Chile,
which is equipped with the first routinely operated multi-color instrument providing Lucky Imaging photometry (\citealt{Skottfelt.2015.A}).
The camera was operated at a 10 Hz rate and lucky exposures were calibrated and tip-tilt corrected as described by \cite{Harpsoe.2012.A}.
The stacked images were used for obtaining photometry with a modified version of DanDIA. 
In addition, the event was observed by the Microlensing Follow Up Network ($\mu$FUN) 0.35m telescope at the Possum Observatory in New Zealand
and by the $\mu$FUN 0.3m Perth Exoplanet Survey Telescope (PEST) 
in Australia, both with a ``clear'' filter. These observations densely cover the first bump.
Finally, the event was also observed by the $\mu$FUN 1.3m SMARTS telescope at CTIO, with ANDICAM, giving simultaneous $I$ band and $H$ band measurements
(a few additional $V$ band observations were taken). This multi-filter imaging was important for the source characterization,
complementing the OGLE observations.
All $\mu$FUN data were reduced using DoPhot \citep{Schechter.1993.A}.

In summary, the results reported here depend overwhelmingly on follow-up data, both to cover the anomaly and for color information of the source,
which makes it possible to both detect and characterize the lens companion.

\subsection{{\it Spitzer} observations}

The $Spitzer$ team modeled and predicted the evolution of all ongoing microlensing events, on a daily-basis,
prior to and during the six weeks of the 2015 campaign.
The team realized the high-magnification nature of OGLE-2015-BLG-1319, and thus its potential high planet sensitivity,
from preliminary OGLE and MOA data, when the event was only 1.4 magnitudes brighter than the baseline
(ultimately getting 4.4 mag brighter than base --- see Figure~\ref{fig:lc}).
\cite{Yee.2015.A} describe the $Spitzer$ selection criteria and observing strategy for such events.
Since the event was in a low-cadence OGLE field and was not observed at all by KMTNet,
it failed to meet criterion B2, and thus could not be selected objectively.
In addition, the estimated flux in $Spitzer$'s $L$ band on the date of first possible observations (six days later on HJD'=7205) and beyond,
assuming the ground-based light curve,
was too faint, $L_{\rm eff}>15.9$ (see definition in \citealt{Yee.2015.A}).
Nevertheless, the team selected the event subjectively due to its predicted high magnification,
immediately announced it as a $Spitzer$ target, and urged the follow-up teams to monitor the event in order to have high planet sensitivity.

OGLE-2015-BLG-1319 was observed by $Spitzer$ on the final two weeks of the 2015 campaign (July 3--19),
with a cadence of 1--2 times per day, and with each epoch composed of six 30-second dithered exposures.
The observations covered the peak of the event as seen from $Spitzer$,
due to the microlens parallax,
when the ground-based light curve was already almost back to baseline. 
The event was indeed faint in $Spitzer$ images, reaching $L_{\rm eff}=15.9$, fainter than the assumed sensitivity limit.
However, the data were reduced using the new algorithm for $Spitzer$ photometry in crowded fields (\citealt{Calchi.2015.B}),
resulting in the required high precision.
It is important to note that while the peak $L_{\rm eff}$ was similar to the estimation based on the ground-based light curve,
it occurred at a later time and with lower magnification than predicted. However, since the source color $(I-L)_{S,0}=1.35$ (see Section~\ref{sec:cmd} below)
was redder than the default assumption for dwarfs of 0.8 by \cite{Yee.2015.A}, coincidentally, the final brightness was similar.

\subsection{{\it Swift} observations}

Based on preliminary estimates for the peak magnification of $A\sim1000$ and first hints of anomaly over peak,
RS requested ToO observations of OGLE-2015-BLG-1319 with $Swift$.
Observations were approved and carried out on June 27, 28, and 29 using the UVOT camera with the $V$ filter.
Each of the three $Swift$ epochs is composed of a sequence of 3 exposures (200s, 200s and 90s).
For the first epoch we use each image separately, while for each of the other two epochs we use a co-added image.
The photometry was extracted using DoPhot. 

\begin{figure}
\plotone{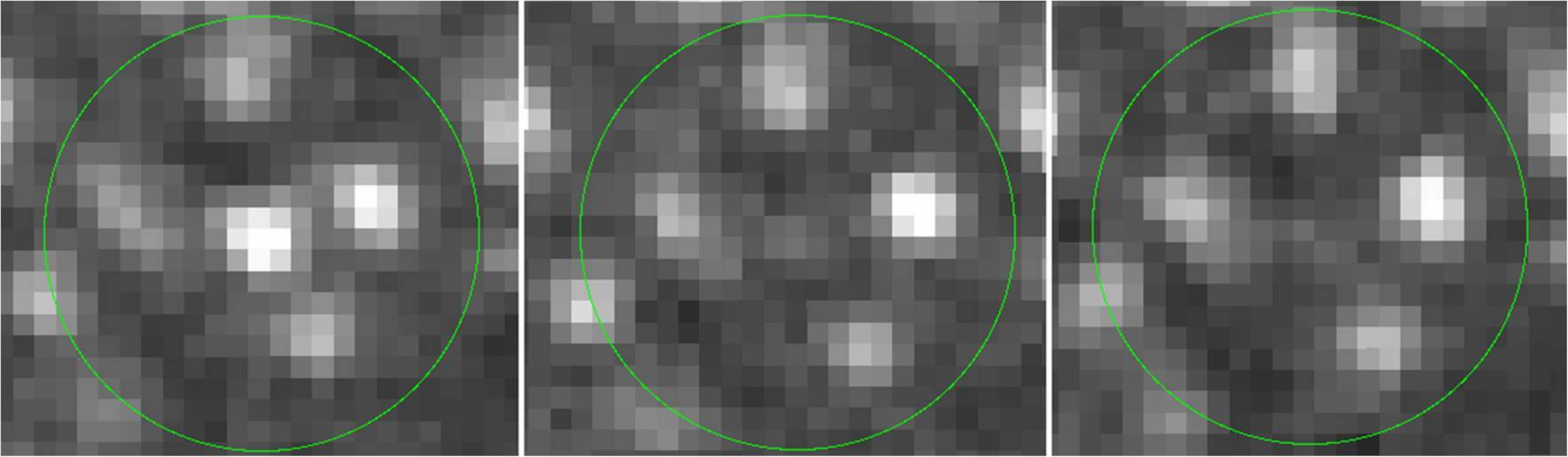}
\caption{Co-added UVOT images for the three epochs when the event was observed with $Swift$ (on June 27, 28, 29, from left to right).
This is the first time $Swift$ observed a microlensing event while it was ongoing, with magnifications of 530, 90, and 50.
The event is clearly seen and detected only on the first epoch, when it was $V=16.6$.
}
\label{fig:swift_im}
\end{figure}

This is the first microlensing event observed with $Swift$.
Figure~\ref{fig:swift_im} shows the co-added image for each epoch.
The event is clearly seen in the first epoch, when the event was highly magnified as seen from Earth,
and is marginally detected in the other two epochs,
when the event was almost 2 magnitudes fainter.
While these observations cannot set significant constraints on the microlensing model,
in particular on $\pi_\e$, they allow us to study the feasibility of $Swift$ observations
for microlensing events, as we do below in Section~\ref{sec:swift}.

{\section{Light Curve Analysis}
\label{sec:anal}}

{\subsection{Ground-only microlensing model}
\label{sec:gblc}}

\begin{figure}
\plotone{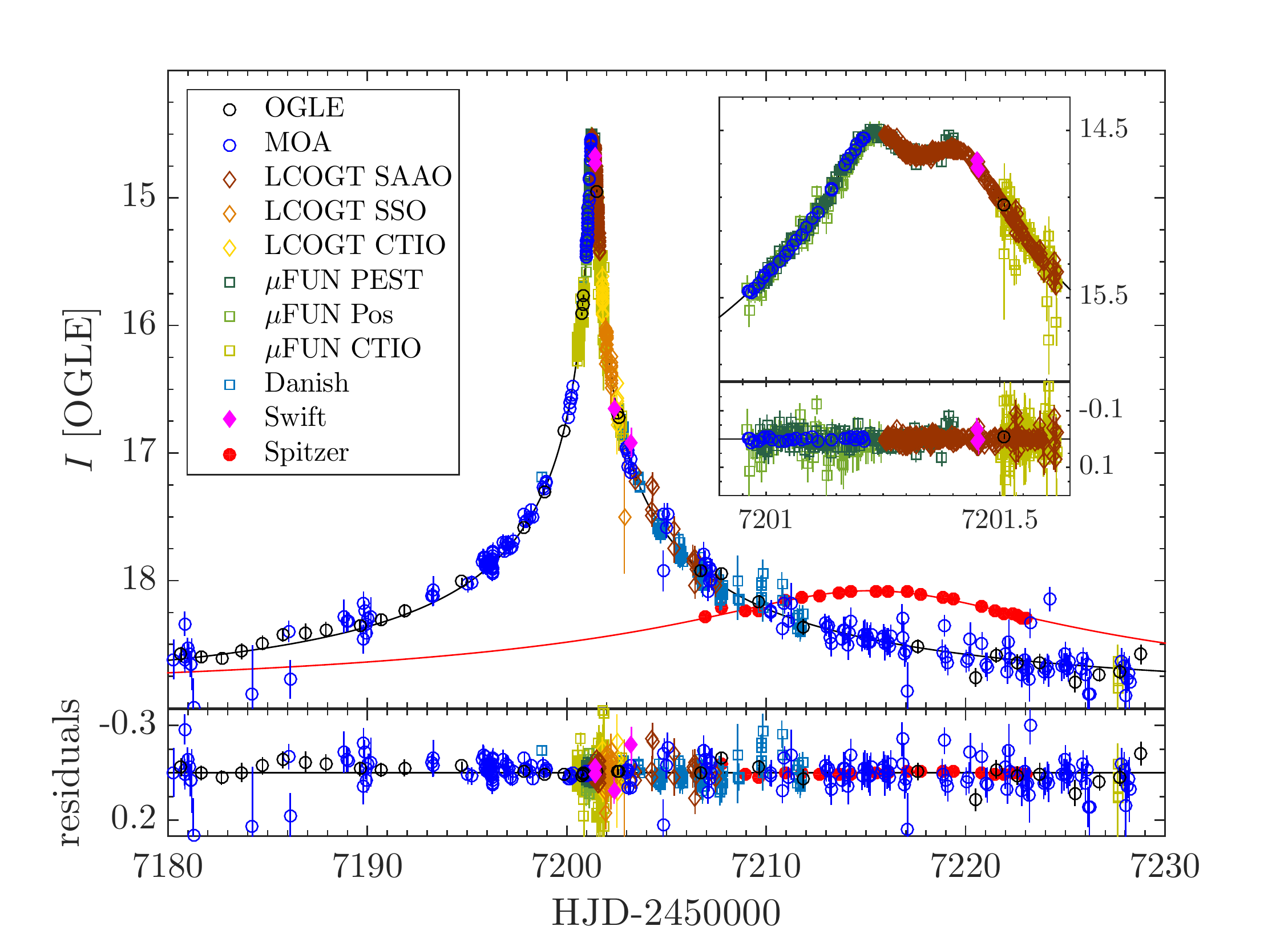}
\caption{Light curve of OGLE-2015-BLG-1319 with data from $Spitzer$ (red), $Swift$ (magenta),
and various ground-based observatories (see interior figure labels).
All observations are aligned to the OGLE magnitude scale, such that equal ``magnitude''
reflects equal magnification.
The inset shows the anomalous region over the peak of the event, revealing the presence of the companion BD.
The clear offset, both in time and magnification, of the $Spitzer$ data with respect
to the ground data allows us to measure the microlens parallax $\pi_\e$.
}
\label{fig:lc}
\end{figure}

\begin{figure}
\plotone{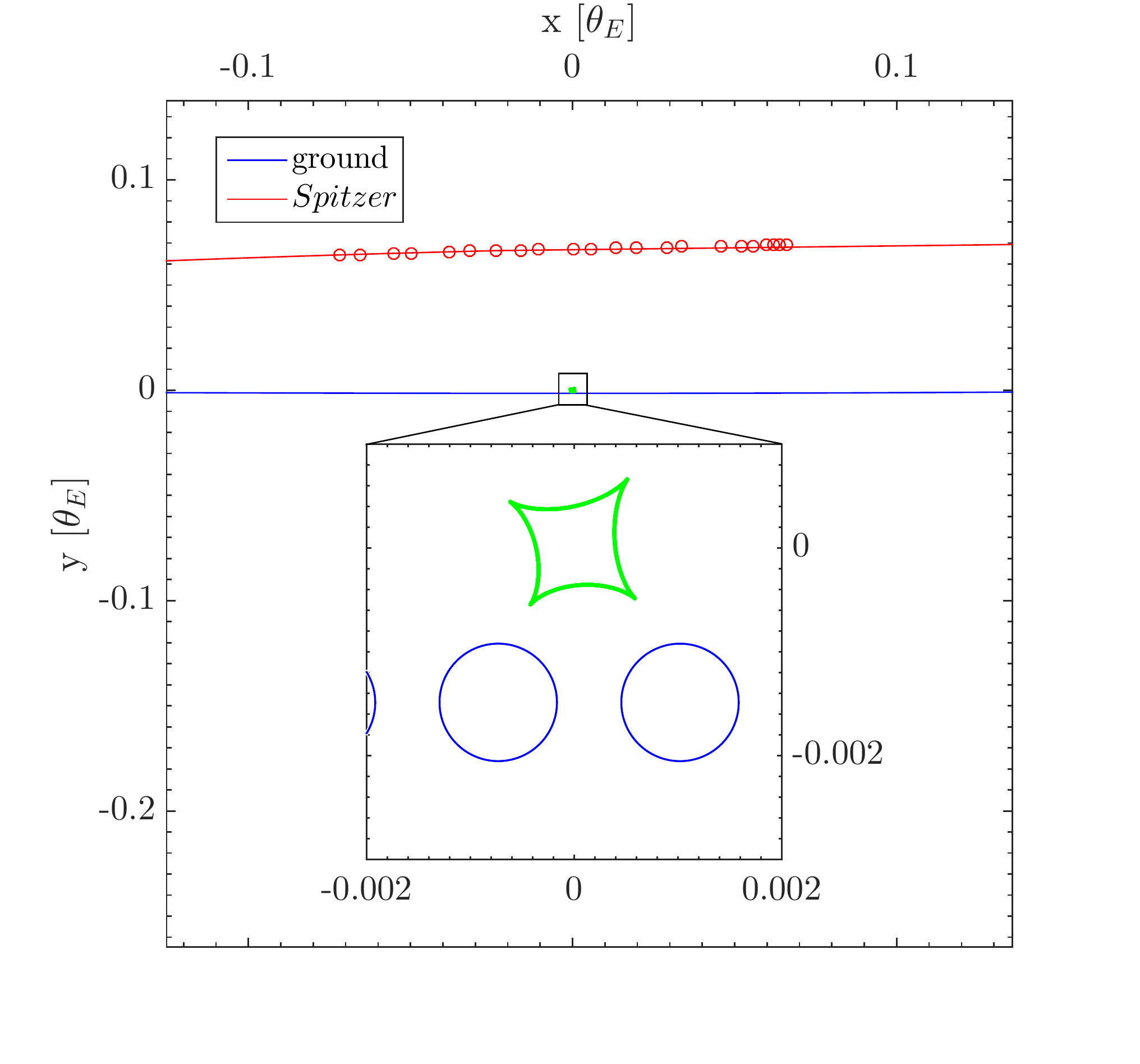}
\caption{The source trajectory as seen from ground (blue) and from $Spitzer$ (red) relative to the central caustic (green),
	 for the wide -/+ configuration. 
	 The coordinate system is defined such that the x-axis is parallel to the source trajectory as seen from the ground at $t_0$. 
	 The red circles represents the source position at the times of $Spitzer$ observations.
	 The inset is a zoomed-in version showing the source angular size (blue circles) for two (arbitrary) times, as seen from the ground.
	 For solutions with $u_0>0$, the caustic structure and the ground trajectory will be (approximately) mirrored,
	 and for solutions with $\pi_{\rm E,N}<0$ the $Spitzer$ trajectory will be (approximately) mirrored, in both cases around the y-axis.}
\label{fig:caustics}
\end{figure}

The light curve of the event, shown in Figure~\ref{fig:lc}, has one ``double-bump'' anomaly over its peak,
while the remainder follows a standard point-lens high-magnification profile.
These features suggest that the source passes near a central caustic with two possible topologies.
Either the source approached close to two cusps of a binary lens system (see Figure~\ref{fig:caustics}),
or three cusps of a planetary system, on the opposite side of the planet. 
\cite{Han.2008.A} studied the planet/binary degeneracy in such double-bump high-magnification events
and showed that if the source passes close enough to the cusps, the light curve will have a characteristic feature distinguishing between the two solutions:
the planetary model will show a flattening between the two bumps while in the binary model it will have a concave shape.
The smooth curved interval between the two bumps seen in OGLE-2015-BLG-1319 thus clearly favors the binary solution.

A standard binary-lens microlensing model requires seven parameters to calculate the magnification as a function of time, $A(t)$.
In addition to the point-lens parameters, $(t_0,u_0,t_\e)$, and the scaled finite source size,
$\rho= \theta_*/\theta_{\rm E}$ (where $\theta_*$ is the angular source size), the companion introduces three parameters.
These are the mass ratio between the companion and the primary, $q$, their scaled, instantaneous projected separation in units of the angular Einstein radius, $s$,  
and an angle, $\alpha$, measured counter-clockwise from the source trajectory to the companion in the lens plane.
For a given model geometry, two flux parameters are assigned for each dataset, $i$, accounting for the source flux, which is being magnified, $f_{s,i}$,
and any additional blend flux, $f_{b,i}$:
\begin{equation}
f_{i}(t) = f_{s,i}A(t) + f_{b,i}.
\label{eqn:fluxpre}
\end{equation}

A Markov-chain Monte-Carlo (MCMC) search of parameter space, with a grid of initial angles of $0^{\circ}<\alpha<360^{\circ}$,
for both the planetary and binary configurations was used to find the best fit model. For the initial modeling we use only $u_0>0$ and $s>1$ to avoid 
possible degenerate solutions, which we address and consider later.
For each set of trial parameters, we fit points far from the two bumps using the hexadecapole, quadrupole,
or monopole approximations \citep{Pejcha.2009.A,Gould.2008.A}, while for points near and during the anomaly we use contour integration \citep{Gould.1997.A}.
For the finite source size we assume a limb-darkened profile with a linear coefficients of $u(V,I,H) =[0.782, 0.607, 0.425]$ \citep{Claret.2000.A},
based on the source type derived below in Section~\ref{sec:cmd}.
We find, as expected, that the binary model is favored over the planetary model by $\Delta\chi^2\sim300$.
\cite{Dominik.1999.A} and \cite{Bozza.2000.A} predicted that for Chang-Refsdal lenses there is an  $s\leftrightarrow s^{-1}$ wide/close degeneracy.
We search and find a close solution ($s\simeq0.08$) that is degenerate with the wide one ($s\simeq14$).

The mass ratio we find is $q\simeq0.08$, suggestive of a low-mass stellar companion.
The event was highly magnified ($A_{\rm max}\simeq725$) and had a long timescale of $t_\e\approx100$d.
For such a timescale, the orbital microlens parallax due the orbital motion of Earth can be detected, 
but since it was heavily blended ($f_s/f_b\simeq0.1$) it appeared magnified for only $\sim40$ days.
Including orbital parallax improves the fit by only $\Delta\chi^2=6$, which is within our systematic uncertainty range (see \ref{sec:close_wide}),
and thus we do not consider it as a detection.
We next include $Spitzer$ data and fully constrain the microlens parallax.

{\subsection{Satellite microlens parallax}
\label{sec:parms+pie}}

Observations by two fixed observers introduce a four-fold degeneracy, as discussed above.
If the microlens parallax can be detected separately in one or both of the observed light curves, due to the orbital motion of the observer,
the degeneracy can be completely removed.
Since the orbital parallax is only marginally detected in the ground-based light curve of OGLE-2015-BLG-1319,
when including $Spitzer$, we re-run the MCMC process with all four possibilities for both the wide and close configurations.
In addition, we include a constraint on the $Spitzer$ source flux, $f_{s,Spitzer}$, derived from color-color regression (see Section~\ref{sec:cmd} below). 

The results find that indeed the eight possible solutions, the four-fold satellite degeneracy for both the wide and close configurations, are fully degenerate.
The microlens parallax components are roughly the same for all solutions since the impact parameter, $u_0$, as seen from Earth, is very close to zero
(see \citealt{Gould.2012.A}).
The magnitude of the microlens parallax is $\pi_\e\simeq0.12$, with 4-8\% uncertainty.
As a check, we run the chains without the $Spitzer$ flux constraint. We find that the median $I_{\rm OGLE}-L_{Spitzer}$ source color is similar to the one
derived from regression, but the source-color distribution from the MCMC is wider than the constraint uncertainty.
This would imply a 10-20\% uncertainty on the microlens parallax magnitude in the absence of the flux constraint. 

Table~\ref{tab:model} summarizes the derived model parameters and their uncertainties for the eight degenerate solutions.
While in this case, the satellite degeneracy has no importance for the physical interpretation of the lensing system,
the wide/close degeneracy suggests two significantly different orbital periods for the companion.
We next try to resolve this degeneracy.

{\subsection{Close/Wide degeneracy}
\label{sec:close_wide}}

A binary lensing system with projected separation significantly different from the angular Einstein radius can be approximated
as a circularly symmetric system with a weak perturbation potential. 
\cite{An.2005.A} studied this symmetry and its implication on binary-lens
microlensing light curves and found an analytic form to convert between a set of $(s_{\rm close},q_{\rm close})$ to their degenerate pair $(s_{\rm wide},q_{\rm wide})$.
The degeneracy is more severe when $s\ll1$ or $s\gg1$, i.e., far from the resonant caustic topology.
The projected separation of OGLE-2015-BLG-1319 is $s\simeq0.08$ (or $s\simeq14$), which is securely in the highly degenerate regime.

A possible way to distinguish between the two degenerate solutions is by detecting the projected orbital motion of the companion.
This requires two additional parameters representing the evolution of the companion's position (angle and separation) during the event, 
$d\alpha/dt$ and $ds/dt$.
For a single observer these are commonly degenerate with the microlens parallax, since both can have similar signatures on the light curve.
However, when considering two or more observers, as in our case, the microlens parallax information comes from a completely different and independent measurement,
and thus they can easily be disentangled.
The models of the wide and close configurations, when including orbital motion, will usually still be degenerate (or very close to it), thus neither can be favored
by goodness-of-fit tests. However, they can be distinguished by energy considerations.
Each solution implies a certain ratio of the projected kinetic to potential energy \citep{Dong.2009.B},
\begin{equation}
\beta = {\rm \bigg(\frac{KE}{PE}\bigg)_\perp} = \frac{v_\perp^2 r_\perp}{2GM} = \frac{\kappa M_\odot {\rm yr}^2}{8 \pi^2}\frac{\pi_\e s^3 \gamma^2}{\theta_\e(\pi_\e+\pi_S/\theta_\e)^3},
\label{eqn:energy}
\end{equation}
where $\gamma^2=(ds/dt/s)^2+(d\alpha/dt)^2$.
The typical ratio is $\beta\sim \mathcal{O}(0.4)$.
The main observables that are different between the wide and close models, in the limit of low $q$, are $(s,ds/dt,d\alpha/dt)$
($\theta_\e$ will also be different by $\sim\sqrt{1+q}$).
Since the rate of position change should be very similar,
giving approximately the same $\gamma^2$, the dominant difference will be due to $s$.
The strong dependence on the projected separation and the $s\leftrightarrow s^{-1}$ nature of the wide/close degeneracy
suggest $\beta_{\rm wide}/\beta_{\rm close}\simeq s^6$. Thus, if the close solution has a typical value,
the wide solution will give $\beta\gg 1$, an unphysical solution. Conversely, if the wide solution has a typical energy ratio, then the close solution will have $\beta\ll 1$,
which has a negligible probability (the probability distribution follows $\beta^2$).

We include the possibility of orbital motion in the eight solutions for OGLE-2015-BLG-1319.
For the wide solutions the fit is improved by $\Delta\chi^2=17$ and for the close solutions by $\Delta\chi^2=6$.
The projected energy ratio for the wide solutions is $\beta_{\rm wide}=3.6\times10^5$ and for the close solutions is $\beta_{\rm close}=0.7$.
Thus, while the wide configuration gives a better fit, it is ruled out by physics and therefore shows that systematic errors are possible at the
$\Delta\chi^2=11$ level. Hence, the improvement for the close solution is not believable. Even if we ignore systematics, the $\Delta\chi^2=6$ improvement
is too weak to reliably claim a detection.
We conclude that orbital motion cannot be reliably detected and so the wide/close degeneracy remains.

{\section{Physical properties}
\label{sec:physical}}

The mass and distance of the lensing system can be derived from the microlens parallax and the angular Einstein radius,
\begin{equation}
M = \frac{\theta_\e}{\kappa \pi_\e};
\qquad
\pi_\rel = \pi_\e\theta_\e.
\label{eqn:meqn}
\end{equation}
These allow us to translate the mass ratio and the scaled projected separation between the two companions to absolute physical values.
While $\pi_\e$ is a direct observable, the angular Einstein radius $\theta_\e$ is derived from the scaled finite source size (found from the light curve model) 
and the angular size of the source (found using the CMD) by $\theta_{\rm E}=\theta_*/\rho$.

{\subsection{Color-magnitude diagram}
\label{sec:cmd}}

\begin{figure}
\plotone{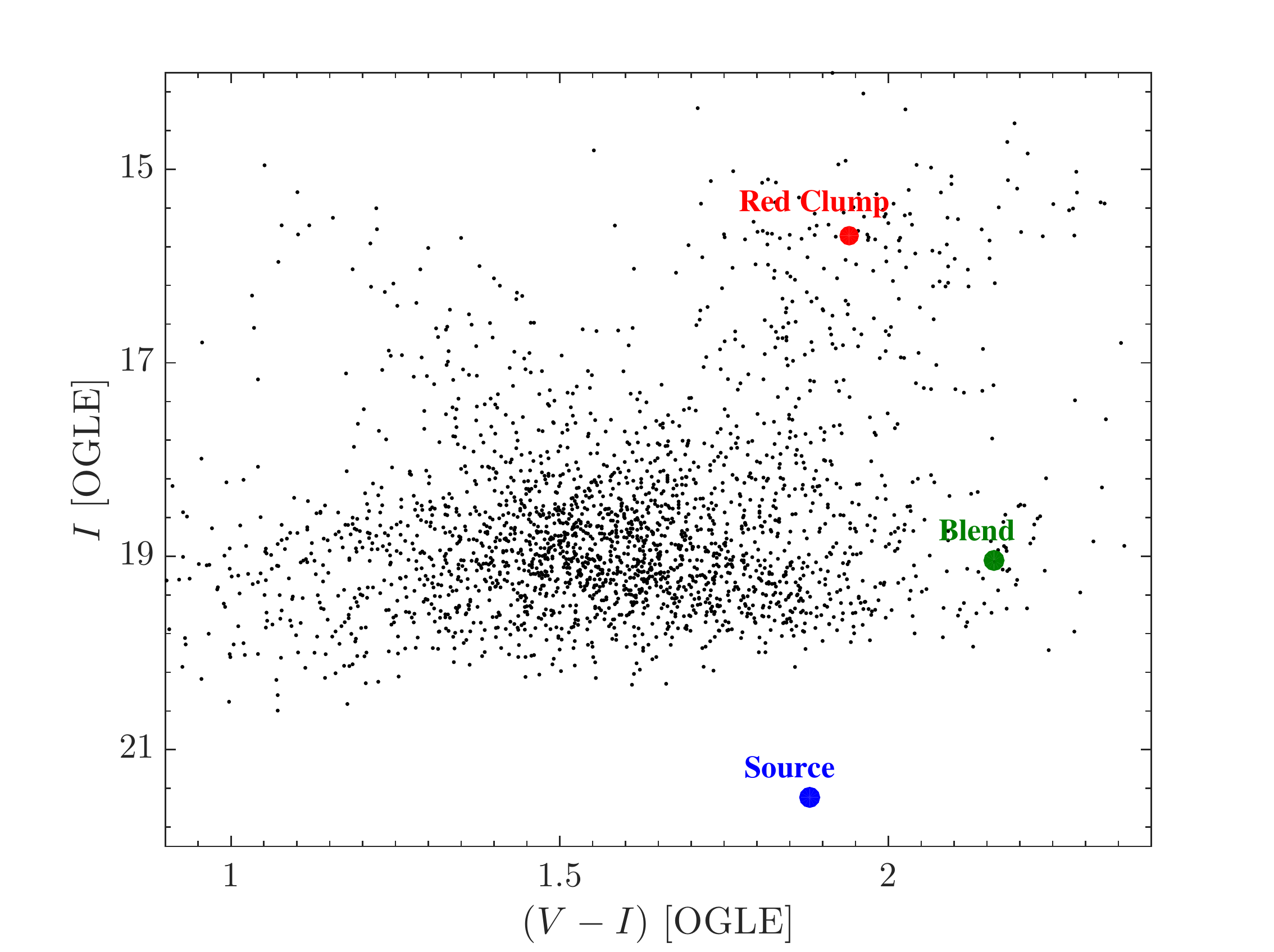}
\caption{OGLE instrumental CMD of stars within 90$''$ of the event's position.
The offset between the red clump centroid (red) and the source star (blue) allow us to derive the source angular radius $\theta_*$.
The total ``blend light'' (green), which is composed of the light from the lens and additional unrelated stars in the OGLE PSF, is also marked.
From the derived mass and distance of the lens star we find that the lens is not the dominant object of the blend. 
}
\label{fig:cmd}
\end{figure}

The source properties can be derived from its position on a CMD.
We construct a CMD of objects within 90$''$ of the event's position (Figure~\ref{fig:cmd}), using OGLE instrumental $V$-band and $I$-band magnitudes. 
We estimate the centroid of the ``red giant clump'' (RGC) to be at 
$(V-I,I)_{\rm cl, ogle}=(1.94,15.68)$ and compare it to the intrinsic centroid of $(V-I,I)_{\rm cl,0}=(1.06,14.51)$ derived by \cite{Bensby.2013.A} and \cite{Nataf.2013.A}
for the Galactic coordinates of the event, $(l,b)=(-1.7,-4.0)$.
The source baseline OGLE $I$-band magnitude as inferred from the microlensing model is $I_{s, {\rm ogle}}=21.50\pm 0.06$,
and assuming it is behind the same dust column as the red clump, its intrinsic magnitude is $I_{s,0}=20.33\pm 0.06$.
This is one of the faintest sources ever reported in microlensing.

It is important to determine the source $(V-I)$ color for two reasons.
First, this quantity enters into the measurement of the angular source size
$\theta_*$. Second, it is needed to estimate the $I-L_{Spitzer}$ source color, via
a $VIL$ color-color diagram derived from nearby field stars.

A standard way to determine the instrumental $(V-I)_s$ color is from
regression of $V$ versus $I$ flux as the source magnification changes.
We apply this technique to OGLE data and find $(V-I)_{s, {\rm ogle}}=1.87\pm0.07$, from
which we derive $(V-I)_{s,0}=1.00\pm 0.07$ by correcting to the clump offset
found above.
Note that we must also account for the fact that the OGLE instrumental
$(V-I)$ is a factor of 1.09 larger than standard.  This
measurement suffers from two separate potential problems.  First,
the error is relatively large.  Second, almost all the information
in the regression comes from a single, significantly magnified $V$
point and so could be subject to systematic errors.
Therefore, we also measure $(V-I)_{s,0}$ using a second method.
We apply regression to determine the instrumental $(I-H)_{s, {\rm ctio}} = 0.67\pm0.01$
from $I$ and $H$ data taken simultaneously at SMARTS CTIO.
We then used a $VIH$ color-color relation derived from all stars in our field
to infer an instrumental $(V-I)_{s,\rm ctio} = 1.64 \pm 0.03$.  Then, comparing
to the instrumental clump $(V-I)_{\rm cl, ctio} = 1.69$, we derive
$(V-I)_{s,0}=1.01\pm 0.03$. 

Combining the two measurements yields $(V-I)_{s,0}=1.01\pm 0.03$.
Using standard color-color relations (\citealt{Bessell.1988.A}) and the relation between angular source size and surface brightness, derived by \citep{Kervella.2004.A},
we find $\theta_*=0.38\pm 0.02\,\muas$.

Before applying the $VIL$ color-color relation (using
OGLE-$IV$ and $Spitzer-L$) to derive the
$(I-L)_s$ color we convert the $(V-I)_{s,0}$ color back to the
OGLE system, i.e., $(V-I)_{s,\rm ogle}=1.88\pm 0.03$.
We then use red giant branch stars ($I_{\rm ogle}<18\,;\, 1.7<(V-I)_{\rm ogle}<2.3$), which are a good representation of the bulge star population,
to derive the color-color relation and find the source instrumental color, $(I_{\rm ogle}-L_{Spitzer})_s=1.54\pm 0.07$.
As a check, we also fit with a larger set including most stars ($I_{\rm ogle}<20\,;\, 1<(V-I)_{\rm ogle}<3$) and find a compatible color $1.51\pm 0.02$.

Any light from the lensing system is superposed on the microlensing event and is a part of the blending flux.
Therefore, the blend flux sets an upper limit on the lens flux, giving an additional constraint to the physical properties of the lensing system.
The high blending fraction in OGLE-2015-BLG-1319 suggests that the lens star might dominate the blended flux.
If so, it can have an additional application. For example, one can measure the radial velocity of the lens star (\citealt{Boisse.2015.A,Yee.2016.A}),
which can be a complementary way to break the wide/close degeneracy.
Subtracting the source color and magnitude from the total baseline fluxes gives $(V-I,I)_{\rm blend, ogle} = (2.16,19.04)$.
If we assume that the blend is behind all the dust, which for the direction of the event means it is at $\gtrsim3$ kpc \citep{Green.2015.A},
we find $(V-I,I)_{{\rm blend},0} = (1.26,17.87)$.
We next derive the physical properties of the lensing system and show that the lens cannot be the dominant source of the blend light.

{\subsection{Another brown dwarf in the desert?}
\label{sec:BDs}}

Table~\ref{tab:physical} summarizes the physical properties derived for each of the eight degenerate solutions.
The results span a relatively narrow range for each of the properties, but one that is still wider than the uncertainties of each of the solutions,
therefore we discuss the full 1$\sigma$ range of all results.

The angular Einstein radius we find is $\theta_\e=$0.54--0.78 mas.
The relative proper motion between the source and the lens is $\mu_{\rm hel}=$1.8--2 mas/year.
This is smaller than what is typical for disk lenses, and moreover the direction is also peculiar.
Converting the relative proper motion to the local system of rest (LSR), we find two solutions (corresponding to $\pm\pi_{\rm E,N}$)
$\mu_{\rm LSR}(l,b)=(-1.9,0.6)$ mas/yr or $\mu_{\rm LSR}(l,b)=(-0.1,1.8)$ mas/yr.
Since the typical value is $\mu_{\rm LSR}(l,b)=(6.5,0.0)\pm(2.9,2.8)$ mas/yr, the probabilities for the solutions are only 2\% and 6\%, respectively. 

The host lens is a 0.44--0.8 $M_\odot$ star and the system is at a distance of 4.6-5.1 kpc.
These suggest a K dwarf host fainter than $I_{l,0}>18.4$.
Therefore, the lens star is not the dominant source of blend light.
The companion is a 30--55 $M_J$ BD, with two possible solutions for the projected separation due to the wide/close degeneracy,
of either 0.23--0.28 AU or 40-52 AU. The close solution suggests that the BD is inside the the ``brown-dwarf desert'' (e.g. \citealt{Grether.2006.A}).
We discuss this possibility below in Section~\ref{sec:discussion}. 

{\section{$Swift$ feasibility of microlens parallax measurements}
\label{sec:swift}}

The microlens parallax sensitivity of a satellite depends on its projected separation from Earth, the photometric precision, and the underlying event microlens parameters,
mainly $u_0$, $t_\e$ and $\pi_\e$.
In the case of a low-Earth-orbit satellite such as $Swift$ ($\sim$600 km from the surface of the Earth),
the separation might be too small to detect the parallax signal in a typical event \citep{Honma.1999.A,Gould.2013.A}.
Recently, \cite{Mogavero.2016.A} showed that a low-Earth-orbit satellite can be used to discover free-floating planets down to the mass of the Earth,
i.e., with short timescales of $t_\e<1$ days, for impact parameters of $u_0<0.1$ (i.e., $A_{max}>10$).
They assumed a continuous 3-minute cadence with a $\sigma_m=0.01$ mag photometric precision at baseline, while $Swift$ has only $\sigma_m=0.1$ mag precision at $V\simeq17.3$ mag
(for 3-minute exposures), which corresponds to RGC bulge stars.
Most microlensing sources are 3-4 magnitudes fainter at baseline.
Even for RGC sources, the magnification needs to be $>$100 times higher (i.e. $u_0<10^{-3}$) in order to compensate for the photometric precision
(though for such giant sources the finite source size starts to limit the maximal magnification for such impact parameters).

\begin{figure}
\plotone{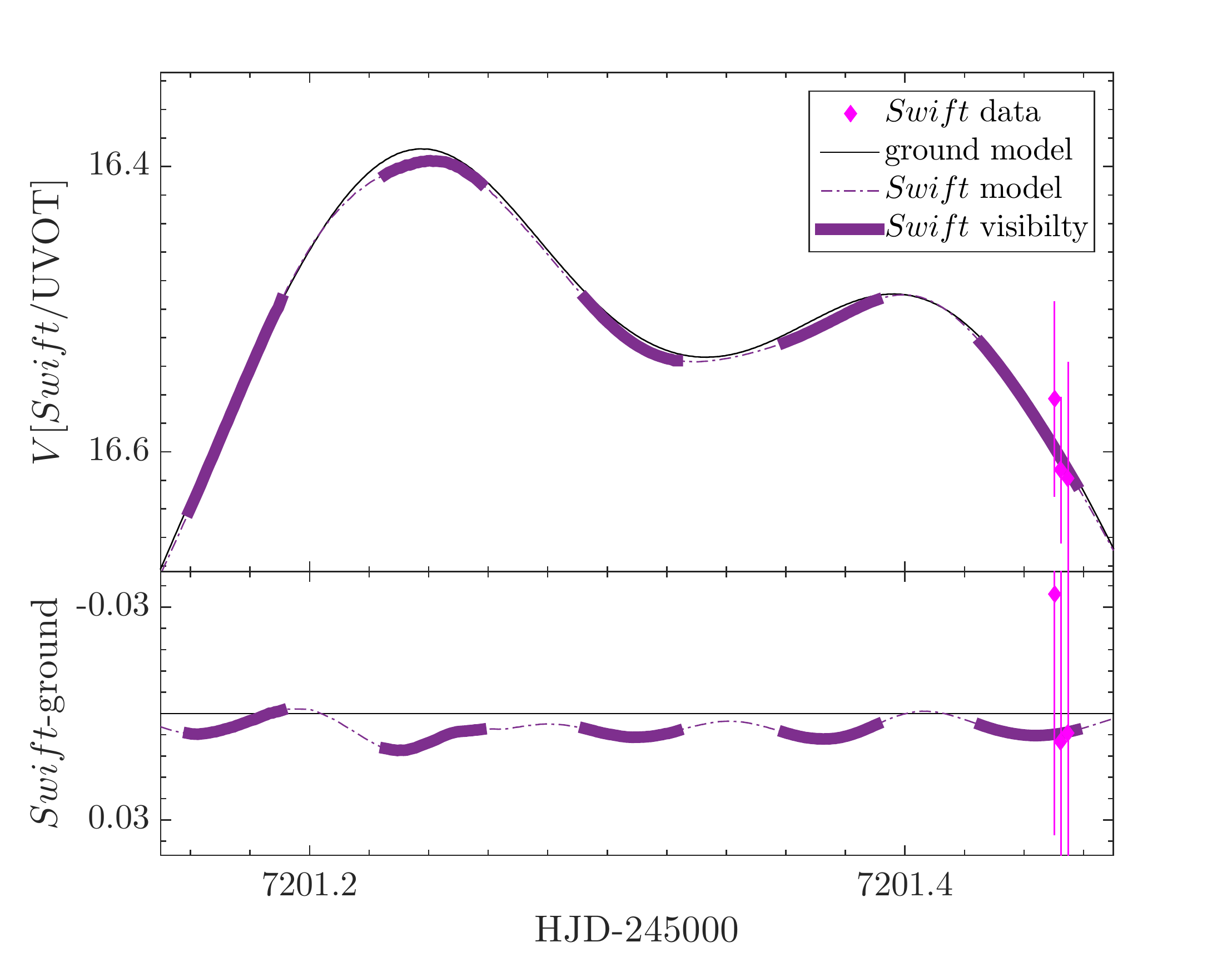}
\caption{The light curve model of OGLE-2015-BLG-1319 as seen from $Swift$ (dash-dot purple line) and from the ground (black line).
The thick lines show the 50-minute windows during which the event was visible from $Swift$, and the magenta diamonds are the three measurements during the first epoch.  
The wave-like structure between the models due to the varying projected separation of $Swift$ allows, in principle, for the detection of the microlens parallax.
However, the maximal difference in the case of OGLE-2015-BLG-1319 was only 0.006 mag (at the first bump). Even if the event were observed at that time,
the $Swift$ precision was not sufficient to constrain the microlens parallax $\pi_\e$.
}
\label{fig:swift_lc}
\end{figure}

In the case of OGLE-2015-BLG-1319, the microlens parallax could not be detected with $Swift$ data.
However, it can be used as a first test case for such measurements. 
In particular, the microlens parallax {\bf was} measured by $Spitzer$,
so we know what the $Swift$ light curve should look like.
Although it was a high magnification event with $A_{\rm max}\sim725$, it was also heavily blended and with a long timescale of $\sim100$ days.
The projected Einstein radius $\tilde{r}_{\rm E}= AU/\pi_\e$ was 7.7 AU, several orders of magnitudes larger than $Swift$ distance from Earth.
The event was bright enough ($V<16.5$) to obtain sufficient photometric precision for only 10 hours, and it was observed for only 10 minutes during that time,
with three consecutive observations.
The field was visible from the satellite for $\sim$50 minutes every orbit of 96 minutes.
Figure~\ref{fig:swift_lc} shows the model of the event as seen from Earth and from $Swift$.
It includes $Swift$ measurements and highlights, in thick lines, the regions where the field was visible from the satellite.
It is clear that $Swift$ could not detect the parallax signal.
The wave-like feature between the two light curves, mainly seen around the first bump, is due to the varying projected separation of $Swift$.
Identifying this feature in future events can be important, allowing for the detection of microlens parallax from relatively short portions of the light curve,
when the precision is sufficiently high.

A study of $Swift$ feasibility for measuring the microlens parallax, extending the work by \cite{Mogavero.2016.A}, requires taking into account the $Swift$ specifications
(photometric precision, bulge visibility from its orbit) as well as more extreme parallax cases. \cite{Mogavero.2016.A} assumed a single (typical) value for $\pi_\rel$ and $\mu$.
Under these assumptions, a given $t_\e$ sets the mass of the lens and the magnitude of $\pi_\e$.
For more extreme values of $\pi_\rel$ and lens mass one can get much larger $\pi_\e$,
to which $Swift$ might be sensitive. In addition, \cite{Mogavero.2016.A} concluded that visibility gaps in the light curves will not significantly alter the parallax sensitivity,
while for high magnification events the effective time is of order of minutes thus it is very sensitive to the $\sim$45 minutes visibility gaps.

\begin{figure}
\plotone{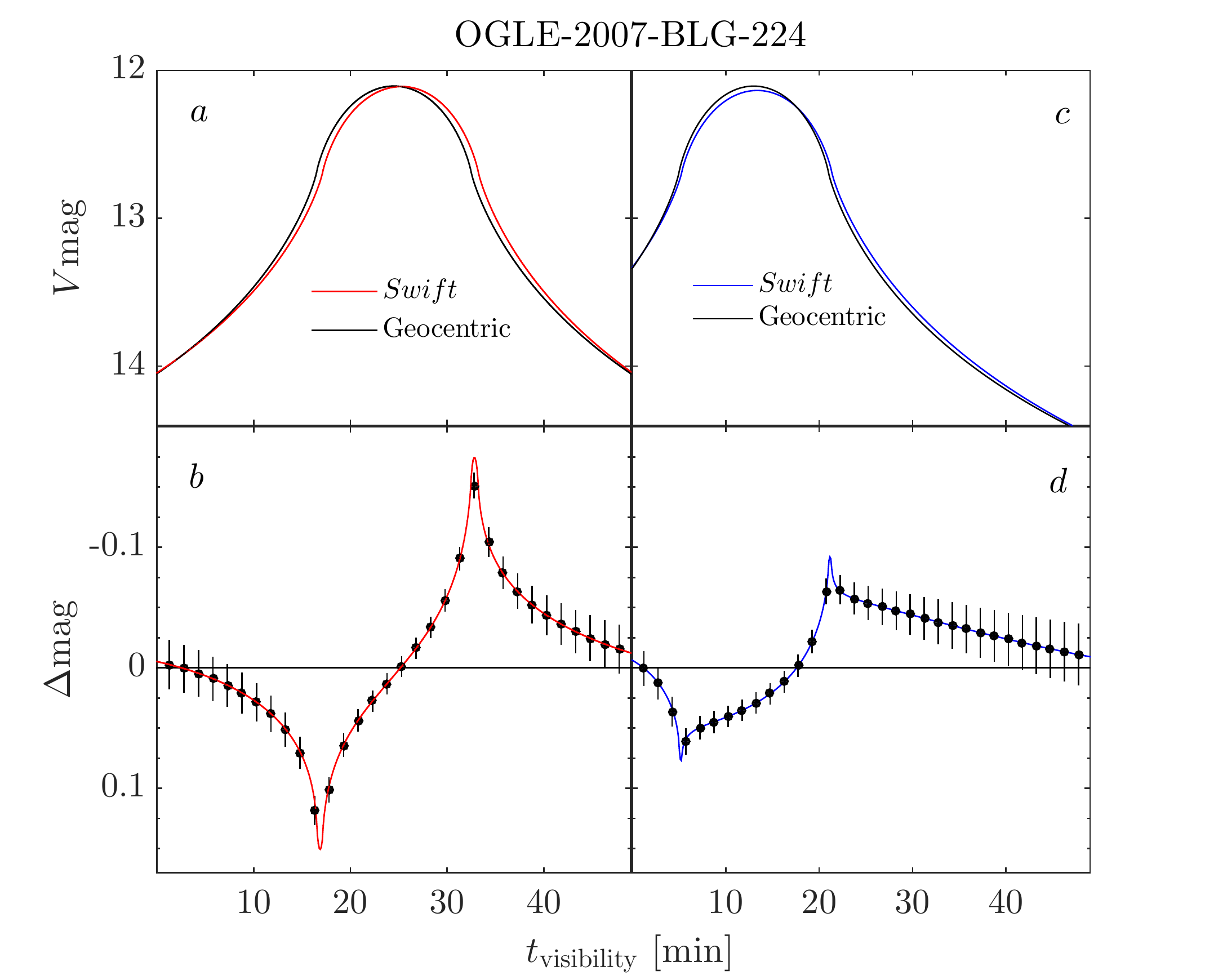}
\caption{Extreme microlensing event OGLE-2007-BLG-224, for which the lens star was a nearby ($\sim$500 pc) single BD, as hypothetically seen by a geocentric observer and from $Swift$.
Panel $a$ shows the peak as seen during the 50-minute $Swift$ visibility window if it was centered around the geocentric $t_0$, while in panel $c$ the geocentric $t_0$ is 13 minutes earlier.
Panels $b$ and $d$ show the magnitude difference between the $Swift$ and geocentric light curves for these two possibilities, respectively. The difference is clearly larger in panel $b$
showing that the $Swift$ feasibility to measure the microlens parallax is very sensitive to the exact time of peak during the visibility window.
The black circles are hypothetical 90 sec exposure data points with the expected $Swift$ precision as their errors. In both cases the microlens parallax is detected, with $\chi^2=931$ for panel $b$ and $\chi^2=670$ for panel $d$.
Since the variations are on timescales of minutes, especially near the maximal differences, the signal is marginalized over the exposure time. This can be seen by the data point at the peak of panel $b$.
For 180 sec exposures, while the photometric precision is better, the overall $\chi^2$ is smaller by 30.
}
\label{fig:swift_ob07224}
\end{figure}

To show $Swift$'s potential for measuring microlens parallaxes we use the extreme microlensing event OGLE-2007-BLG-224 (\citealt{Gould.2009.B})
for which terrestrial parallax was measured. The lens star was a 58$\pm$4 $M_J$ BD at a distance of 525$\pm$40 pc implying a large microlens parallax of 
$\pi_\e=1.97\pm0.13$. The source baseline magnitude was $V_S=20.58$ and the event reached a high magnification of $A=2452$ (i.e., $V_{\rm peak}=12.11$ mag).
Figure~\ref{fig:swift_ob07224} shows the peak of the event as it would have hypothetically seen by a geocentric observer and from $Swift$ during its 50-minute visibility window.
The left panels show the two light curves and their difference if the geocentric $t_0$ was at the middle of the $Swift$ visibility window and the right panels
for a geocentric $t_0$ thirteen minutes earlier. The difference is clear in both possibilities but significantly larger if the visibility window happened to be centered on the geocentric $t_0$,
with maximal difference of $\Delta$mag=0.17 (only $\Delta$mag=0.09 for the offset case),
showing the large sensitivity to the exact time of peak during the visibility window.

We sample the $Swift$ light curve with a 90-sec cadence and use UVOT signal-to-noise
tool\footnote{\url{http://www.mssl.ucl.ac.uk/www_astro/uvot/uvot_observing/uvot_tool.html}}
to estimate the photometric precision for each point. The crowded fields of the bulge could potentially result in larger errors. However, we calculated the $V$ magnitude
root-mean-square (RMS) of all stars in our six 200-sec $Swift$ images of OGLE-2015-BLG-1319 and found that the RMS distribution agrees well with the UVOT tool estimates for 200 sec exposures.
(We note that for the bright $V$ magnitudes OGLE-2007-BLG-224 reached, the coincidence loss correction needs to be taken into account).
We then calculate the $\chi^2$ between the 32 $Swift$ points and the geocentric light curve and find $\chi^2=931$ for the centered observations and $\chi^2=670$ for the offset ones,
thus easily detected in both cases. Observations on the adjacent visibility windows would not improve the detection since the photometric precision is $>$20 larger than
the parallax signal at those regions.
The differences between the two light curves vary around the maximal differences with timescales of minutes, and therefore the parallax signal is also sensitive to the exposure time
(see for example the data point at the maximal difference on panel $b$ of Figure~\ref{fig:swift_ob07224}).
To check this we also sampled the light curve with 180-sec exposures and find that the $\chi^2$ is smaller by 30.

The $Swift$ feasibility can extend also to nearby free-floating planets. The microlens parallax magnitude will be bigger than that of BDs
but the peak magnification will be lower, limited by the finite source effects. The duration of the signal over the peak will be similar since it is 
dominated by the source crossing time $t_*=\theta_*/\mu$.
Finally, we note that $Swift$ might be able to detect the parallax signal during caustic crossings (in particular the caustic exits) of bright binary events
(analogous to the original idea presented by \citealt{Honma.1999.A}) which are much more common,
and with real-time light curve modeling can be predicted in advance to trigger $Swift$.
A thorough study of $Swift$ microlens parallax sensitivity will be conducted in Street et al. (in preparation).

{\section{Discussion}
\label{sec:discussion}}

We have presented the detection, via simultaneous observations from ground, $Spitzer$ and $Swift$, of a BD orbiting a K dwarf with two degenerate solutions for the projected separation.   
This BD adds to other microlensing-detected BDs in a variety of physical configurations.
\cite{Han.2016arXiv.A} summarizes the 15 published microlensing events with BDs prior to OGLE-2015-BLG-1319.
This list includes one BD hosting a planet,
ten BDs around main sequence stars (nine around M dwarfs and one around G-K dwarf), two binary BD systems, and two isolated BDs.
These were discovered through different surveys with different detection efficiencies, 
making it difficult to derive a statistical conclusion from them (though see \citealt{Ranc.2015.A}).
In addition, more than half of the published BDs with companions were discovered due to central caustic anomalies. Therefore, 
the majority of these suffer from the wide/close degeneracy, which leads to a degeneracy in their derived projected separation.
Nevertheless, the accumulation of detections suggests that BDs around main-sequence stars are not rare at separations of 0.5--20 AU, where microlensing is sensitive
(this range is larger than for exoplanets due to higher detection sensitivity).
This is in contrast to estimates through other techniques, such as radial velocity and transit, who find that BDs are rare ($<1\%$, \citealt{Grether.2006.A})
at closer separations.
One possible explanation for this difference, as suggested by \cite{Shvartzvald.2016.A}, is the different host stars that are mostly probed by each technique
--- FGK stars by radial velocity and transits versus M stars by microlensing.

The event was part of the 2015 $Spitzer$ campaign and was the first to be observed simultaneously from two space telescopes.
$Kepler$ follows the path started by $Spitzer$ and is now conducting, as campaign 9 of its \textit{K2} mission (\textit{K2}C9),
the first space microlensing survey (\citealt{Gould.2013.B,Henderson.2015arXiv.A}).
The \textit{K2}C9 fields are monitored continuously from ground, both to increase the planet sensitivity
and to enable the measurement of the microlens parallax of all events in those fields.  
In addition to detecting bound companions (planets, binaries, stellar remnants),
this will be the first opportunity to measure the masses of free-floating planets,
which are identified by their short timescales, of-order 1 day.
Such events cannot be observed from $Spitzer$, which requires target uploads at least 3 days before observations.
Triple-location observations, such as conducted for OGLE-2015-BLG-1319, are planned with $Spitzer$ and $Kepler$, monitoring events in the \textit{K2}C9 fields during the last 10 days of the $Kepler$ campaign,
when both satellites can observe the bulge (see discussion in \citealt{Calchi.2015arXiv.A}).

For the first time, $Swift$ was used to observe a microlensing event and was able to detect the event while it was magnified.
The unique target-of-opportunity override capability of the $Swift$ spacecraft is, by design, ideal for the observation of transient variables of all
kinds and in particular for microlensing.
However, the microlens parallax could not be measured with $Swift$ data for this event. 
From our preliminary study of $Swift$'s ability to detect the microlens parallax signal
we find that it is sensitive to nearby BDs and free-floating planets in high magnification events ($A>1000$).
If the lens in our example of OGLE-2007-BLG-224 was a free-floating super-Jupiter rather than a BD, with all other characteristics of the event identical,
$Swift$ could have significantly detect ($\chi^2=430$) the microlens parallax signal with continuous 90-sec exposures over a 50-minute visibility window centered on the peak of the event.
Such nearby ($\sim$500 pc) massive free-floating planets could be directly imaged with $JWST$ (for ages $\lesssim$1 Gyr), and would possibly allow to study their formation mechanism
and environment.

\acknowledgments
We thank T. Meshkat and R. Patel for fruitful discussions about directly imaging free-floating planets.
Work by YS and CBH was supported by an
appointment to the NASA Postdoctoral Program at the Jet
Propulsion Laboratory, administered by Universities Space Research Association
through a contract with NASA.
This work makes use of observations from the LCOGT network, which includes three SUPAscopes owned by the University of St Andrews.
The RoboNet programme is an LCOGT Key Project using time allocations from the University of St Andrews,
LCOGT and the University of Heidelberg together with time on the Liverpool Telescope through the Science and Technology Facilities Council (STFC), UK.
This research has made use of the LCOGT Archive, which is operated by the California Institute of Technology, under contract with the Las Cumbres Observatory.
The OGLE project has received funding from the National Science Centre,
Poland, grant MAESTRO 2014/14/A/ST9/00121 to AU.
OGLE Team thanks Profs. M. Kubiak and G. Pietrzy{\'n}ski, former
members of the OGLE team, for their contribution to the collection of
the OGLE photometric data over the past years.
TS acknowledges the financial support from the JSPS, JSPS23103002,JSPS24253004 and JSPS26247023.
The MOA project is supported by the grant JSPS25103508 and 23340064.
Work by AG and SC was supported by JPL grant 1500811.
Work by AG and WZ was supported by NSF grant AST-15168.
Work by JCY was performed under contract with the California Institute of Technology 
(Caltech)/Jet Propulsion Laboratory (JPL) funded by NASA
through the Sagan Fellowship Program executed by the
NASA Exoplanet Science Institute.
This work is based in part
on observations made with the {\it Spitzer} Space Telescope,
which is operated by the Jet Propulsion Laboratory, California Institute of Technology under a contract with
NASA.
The $Spitzer$ Team thanks Christopher S.\ Kochanek for graciously trading us his allocated observing time on the CTIO 1.3m during the $Spitzer$ campaign.
We thank the $Swift$ operations team for approving and implementing the Target of Opportunity request.
This publication was made possible by NPRP grant \# X-019-1-006 from the Qatar National Research Fund (a member of Qatar Foundation). 
Work by SM has been supported by the Strategic Priority Research Program ``The Emergence of Cosmological Structures" of the Chinese Academy
of Sciences Grant No. XDB09000000, and by the National Natural Science Foundation of China (NSFC) under grant numbers 11333003 and 11390372.
M.P.G.H. acknowledges support from the Villum Foundation.
NP acknowledges funding by the Portuguese FCT - Foundation for Science and Technology and the European Social Fund (ref: SFRH/BGCT/113686/2015) and the Gemini-Conicyt Fund,
allocated to project \#32120036.
CITEUC is funded by National Funds through FCT - Foundation for Science and Technology (project: UID/Multi/00611/2013) and
FEDER - European Regional Development Fund through COMPETE 2020 –Operational Programme Competitiveness and Internationalisation (project: POCI-01-0145-FEDER-006922).
GD acknowledges Regione Campania for support from POR-FSE Campania 2014-2020.
Based on data collected by MiNDSTEp with the Danish 1.54 m telescope at the ESO La Silla observatory.

Copyright 2016. All rights reserved.

\begin{landscape}
\scriptsize
\begin{table}
\scriptsize
\centering
\caption{Best-fit microlensing model parameters
and their 68\% uncertainty range derived from the MCMC chain density
for the eight degenerate solutions
\label{tab:model}}
\begin{tabular}{l|cccc|cccc}
\tableline\tableline
Parameter	& \multicolumn{4}{c}{Close}	& \multicolumn{4}{c}{Wide}	\\
		& $--$	& $+-$	& $-+$	& $++$	& $--$	& $+-$	& $-+$	& $++$	\\
\tableline\\[-10pt]
$t_0$ [HJD'']	& $1.30439^{+0.00050}_{-0.00060}$	& $1.30413^{+0.00046}_{-0.00047}$	& $1.30440^{+0.00060}_{-0.00049}$	& $1.30422^{+0.00037}_{-0.00053}$	& $1.30472^{+0.00055}_{-0.00041}$	& $1.30458^{+0.00045}_{-0.00042}$	& $1.30480^{+0.00044}_{-0.00034}$	& $1.30476^{+0.00042}_{-0.00041}$	\\[5pt]
$u_0$	& $-0.00161^{+0.00018}_{-0.00016}$	& $0.00152^{+0.00012}_{-0.00013}$	& $-0.00151^{+0.00012}_{-0.00011}$	& $0.00162^{+0.00009}_{-0.00008}$	& $-0.00154^{+0.00010}_{-0.00008}$	& $0.00146^{+0.00009}_{-0.00013}$	& $-0.00155^{+0.00007}_{-0.00008}$	& $0.00157^{+0.00008}_{-0.00011}$	\\[5pt]
$t_{\rm E}$ [d]	& $95.0^{+11.6}_{-8.4}$	& $100.6^{+9.1}_{-7.0}$	& $101.0^{+8.6}_{-6.9}$	& $94.8^{+4.8}_{-5.2}$	& $99.6^{+7.0}_{-4.9}$	& $104.5^{+10.0}_{-6.4}$	& $98.8^{+4.7}_{-4.8}$	& $97.5^{+7.5}_{-4.7}$	\\[5pt]
$\rho$[$10^{-4}$]	& $6.20^{+0.88}_{-0.82}$	& $5.83^{+0.70}_{-0.64}$	& $5.55^{+0.68}_{-0.63}$	& $6.45^{+0.58}_{-0.57}$	& $5.79^{+0.55}_{-0.59}$	& $5.41^{+0.53}_{-0.55}$	& $5.79^{+0.49}_{-0.55}$	& $6.04^{+0.57}_{-0.60}$	\\[5pt]
$\pi_{\rm E,N}$	& $-0.0607^{+0.0064}_{-0.0059}$	& $-0.0595^{+0.0047}_{-0.0047}$	& $0.0487^{+0.0039}_{-0.0039}$	& $0.0499^{+0.0031}_{-0.0030}$	& $-0.0580^{+0.0040}_{-0.0035}$	& $-0.0573^{+0.0049}_{-0.0041}$	& $0.0500^{+0.0030}_{-0.0025}$	& $0.0482^{+0.0030}_{-0.0035}$	\\[5pt]
$\pi_{\rm E,E}$	& $-0.1160^{+0.0126}_{-0.0111}$	& $-0.1096^{+0.0090}_{-0.0082}$	& $-0.1110^{+0.0086}_{-0.0083}$	& $-0.1182^{+0.0058}_{-0.0068}$	& $-0.1106^{+0.0072}_{-0.0058}$	& $-0.1052^{+0.0091}_{-0.0070}$	& $-0.1136^{+0.0054}_{-0.0057}$	& $-0.1149^{+0.0082}_{-0.0059}$	\\[5pt]
$\alpha$ [rad]	& $0.9156^{+0.0048}_{-0.0040}$	& $5.3656^{+0.0028}_{-0.0030}$	& $0.9127^{+0.0041}_{-0.0041}$	& $5.3637^{+0.0044}_{-0.0038}$	& $0.9121^{+0.0023}_{-0.0025}$	& $5.3689^{+0.0023}_{-0.0021}$	& $0.9113^{+0.0022}_{-0.0022}$	& $5.3689^{+0.0023}_{-0.0023}$	\\[5pt]
$s$	& $0.0801^{+0.0057}_{-0.0054}$	& $0.0811^{+0.0033}_{-0.0023}$	& $0.0748^{+0.0063}_{-0.0031}$	& $0.0848^{+0.0030}_{-0.0060}$	& $14.5470^{+0.0209}_{-0.0175}$	& $14.1238^{+0.0260}_{-0.0147}$	& $14.5959^{+0.0090}_{-0.0095}$	& $14.1049^{+0.0240}_{-0.0263}$	\\[5pt]
$q$	& $0.0729^{+0.0022}_{-0.0028}$	& $0.0660^{+0.0042}_{-0.0033}$	& $0.0770^{+0.0127}_{-0.0121}$	& $0.0656^{+0.0117}_{-0.0059}$	& $0.0933^{+0.0057}_{-0.0071}$	& $0.0831^{+0.0063}_{-0.0080}$	& $0.0950^{+0.0057}_{-0.0050}$	& $0.0898^{+0.0049}_{-0.0070}$	\\[5pt]
$f_s/f_b$	& $0.104^{+0.011}_{-0.012}$	& $0.097^{+0.008}_{-0.009}$	& $0.097^{+0.008}_{-0.008}$	& $0.104^{+0.007}_{-0.005}$	& $0.103^{+0.006}_{-0.008}$	& $0.098^{+0.007}_{-0.010}$	& $0.104^{+0.006}_{-0.005}$	& $0.106^{+0.006}_{-0.009}$	\\[5pt]
$\chi^2$	& 2812	& 2818	& 2812	& 2816	& 2812	& 2818	& 2811	& 2816	\\[5pt]
\tableline\tableline
\end{tabular}
\newline
\raggedright{HJD''=HJD-2457200}
\end{table}

\begin{table}
\scriptsize
\centering
\caption{Physical properties of the binary system for the eight degenerate solutions.
\label{tab:physical}}
\begin{tabular}{l|cccc|cccc}
\tableline\tableline
Parameter	& \multicolumn{4}{c}{Close}	& \multicolumn{4}{c}{Wide}	\\
		& $--$	& $+-$	& $-+$	& $++$	& $--$	& $+-$	& $-+$	& $++$	\\
\tableline\\[-10pt]
$M_1$ [$M_\odot$]	& $0.54^{+0.15}_{-0.10}$	& $0.61^{+0.12}_{-0.10}$	& $0.65^{+0.13}_{-0.10}$	& $0.53^{+0.08}_{-0.06}$	& $0.59^{+0.10}_{-0.07}$	& $0.67^{+0.13}_{-0.09}$	& $0.60^{+0.08}_{-0.07}$	& $0.57^{+0.10}_{-0.07}$	\\[5pt]
$M_2$ [$M_J$]	& $40.79^{+12.73}_{-8.70}$	& $42.07^{+6.99}_{-6.17}$	& $53.16^{+8.78}_{-11.57}$	& $36.78^{+6.70}_{-4.82}$	& $57.76^{+5.77}_{-4.46}$	& $57.74^{+5.87}_{-4.67}$	& $59.14^{+5.41}_{-4.08}$	& $53.53^{+5.31}_{-4.19}$	\\[5pt]
$r_\perp$\,\, [$\au$]	& $0.24^{+0.02}_{-0.01}$	& $0.26^{+0.02}_{-0.02}$	& $0.25^{+0.02}_{-0.02}$	& $0.25^{+0.02}_{-0.02}$	& $46.26^{+3.83}_{-3.18}$	& $47.47^{+4.29}_{-3.37}$	& $46.47^{+3.31}_{-2.78}$	& $43.72^{+3.73}_{-2.99}$	\\[5pt]
$D_L$ [kpc]	& $4.89^{+0.14}_{-0.17}$	& $4.86^{+0.14}_{-0.16}$	& $4.83^{+0.16}_{-0.20}$	& $4.99^{+0.13}_{-0.16}$	& $4.85^{+0.13}_{-0.18}$	& $4.80^{+0.15}_{-0.19}$	& $4.84^{+0.13}_{-0.18}$	& $4.93^{+0.13}_{-0.16}$	\\[5pt]
$\theta_E$ [mas]	& $0.61^{+0.09}_{-0.08}$	& $0.65^{+0.08}_{-0.07}$	& $0.68^{+0.09}_{-0.07}$	& $0.59^{+0.06}_{-0.05}$	& $0.66^{+0.07}_{-0.06}$	& $0.70^{+0.08}_{-0.06}$	& $0.66^{+0.07}_{-0.05}$	& $0.63^{+0.07}_{-0.05}$	\\[5pt]
$\mu_{\rm hel}(N,E)[mas/yr]$	& (-1.1,-1.6)	& (-1.1,-1.6)	& (1.0,-1.7)	& (0.9,-1.6)	& (-1.1,-1.6)	& (-1.2,-1.6)	& (1.0,-1.7)	& (0.9,-1.7)	\\[5pt]
$\mu_{\rm LSR}(l,b)[mas/yr]$	& (-1.8,0.6)	& (-1.9,0.6)	& (-0.1,1.8)	& (-0.1,1.7)	& (-1.9,0.6)	& (-1.9,0.6)	& (-0.1,1.8)	& (-0.1,1.7)	\\[5pt]
\tableline\tableline
\end{tabular}
\end{table}
\end{landscape}

\begin{thebibliography}{50}
\expandafter\ifx\csname natexlab\endcsname\relax\def\natexlab#1{#1}\fi

\bibitem[{{An}(2005)}]{An.2005.A}
{An}, J.~H. 2005, \mnras, 356, 1409

\bibitem[{{Bensby} {et~al.}(2013){Bensby}, {Yee}, {Feltzing}, {Johnson},
  {Gould}, {Cohen}, {Asplund}, {Mel{\'e}ndez}, {Lucatello}, {Han}, {Thompson},
  {Gal-Yam}, {Udalski}, {Bennett}, {Bond}, {Kohei}, {Sumi}, {Suzuki}, {Suzuki},
  {Takino}, {Tristram}, {Yamai}, \& {Yonehara}}]{Bensby.2013.A}
{Bensby}, T., {et~al.} 2013, \aap, 549, A147

\bibitem[{{Bessell} \& {Brett}(1988)}]{Bessell.1988.A}
{Bessell}, M.~S., \& {Brett}, J.~M. 1988, \pasp, 100, 1134

\bibitem[{{Boisse} {et~al.}(2015){Boisse}, {Santerne}, {Beaulieu}, {Fakhardji},
  {Santos}, {Figueira}, {Sousa}, \& {Ranc}}]{Boisse.2015.A}
{Boisse}, I., {Santerne}, A., {Beaulieu}, J.-P., {Fakhardji}, W., {Santos},
  N.~C., {Figueira}, P., {Sousa}, S.~G., \& {Ranc}, C. 2015, \aap, 582, L11

\bibitem[{{Bond} {et~al.}(2001){Bond}, {Abe}, {Dodd}, {Hearnshaw}, {Honda},
  {Jugaku}, {Kilmartin}, {Marles}, {Masuda}, {Matsubara}, {Muraki}, {Nakamura},
  {Nankivell}, {Noda}, {Noguchi}, {Ohnishi}, {Rattenbury}, {Reid}, {Saito},
  {Sato}, {Sekiguchi}, {Skuljan}, {Sullivan}, {Sumi}, {Takeuti}, {Watase},
  {Wilkinson}, {Yamada}, {Yanagisawa}, \& {Yock}}]{Bond.2001.A}
{Bond}, I.~A., {et~al.} 2001, \mnras, 327, 868

\bibitem[{{Bozza}(2000)}]{Bozza.2000.A}
{Bozza}, V. 2000, \aap, 355, 423

\bibitem[{{Bozza} {et~al.}(2016){Bozza}, {Shvartzvald}, {Udalski}, {Calchi
  Novati}, {Bond}, {Han}, {Hundertmark}, {Poleski}, {Pawlak}, {Szyma{\'n}ski},
  {Skowron}, {Mr{\'o}z}, {Koz{\l}owski}, {Wyrzykowski}, {Pietrukowicz},
  {Soszy{\'n}ski}, {Ulaczyk}, {OGLE Group}, {and}, {Beichman}, {Bryden},
  {Carey}, {Fausnaugh}, {Gaudi}, {Gould}, {Henderson}, {Pogge}, {Wibking},
  {Yee}, {Zhu}, {Spitzer Team}, {Abe}, {Asakura}, {Barry}, {Bennett},
  {Bhattacharya}, {Donachie}, {Freeman}, {Fukui}, {Hirao}, {Inayama}, {Itow},
  {Koshimoto}, {Li}, {Ling}, {Masuda}, {Matsubara}, {Muraki}, {Nagakane},
  {Nishioka}, {Ohnishi}, {Oyokawa}, {Rattenbury}, {Saito}, {Sharan},
  {Sullivan}, {Sumi}, {Suzuki}, {Tristram}, {Wakiyama}, {Yonehara}, {MOA
  Group}, {Choi}, {Park}, {Jung}, {Shin}, {Albrow}, {Park}, {Kim}, {Lee},
  {Cha}, {Kim}, {Lee}, {KMTNet Group}, {Dominik}, {J{\o}rgensen}, {Andersen},
  {Bramich}, {Burgdorf}, {Ciceri}, {D'Ago}, {Evans}, {Figuera Jaimes}, {Gu},
  {Hinse}, {Kains}, {Kerins}, {Korhonen}, {Kuffmeier}, {Mancini}, {Popovas},
  {Rabus}, {Rahvar}, {Rasmussen}, {Scarpetta}, {Skottfelt}, {Snodgrass},
  {Southworth}, {Surdej}, {Unda-Sanzana}, {von Essen}, {Wang}, {Wertz},
  {MiNDSTEp}, {Maoz}, {Friedmann}, {Kaspi}, \& {Wise Group}}]{Bozza.2016.A}
{Bozza}, V., {et~al.} 2016, \apj, 820, 79

\bibitem[{{Bramich} {et~al.}(2008){Bramich}, {Vidrih}, {Wyrzykowski}, {Munn},
  {Lin}, {Evans}, {Smith}, {Belokurov}, {Gilmore}, {Zucker}, {Hewett},
  {Watkins}, {Faria}, {Fellhauer}, {Miknaitis}, {Bizyaev}, {Ivezi{\'c}},
  {Schneider}, {Snedden}, {Malanushenko}, {Malanushenko}, \&
  {Pan}}]{Bramich.2008.A}
{Bramich}, D.~M., {et~al.} 2008, \mnras, 386, 887

\bibitem[{{Calchi Novati} \& {Scarpetta}(2015)}]{Calchi.2015arXiv.A}
{Calchi Novati}, S., \& {Scarpetta}, G. 2015, ArXiv e-prints

\bibitem[{{Calchi Novati} {et~al.}(2015{\natexlab{a}}){Calchi Novati}, {Gould},
  {Udalski}, {Menzies}, {Bond}, {Shvartzvald}, {Street}, {Hundertmark},
  {Beichman}, {Yee}, {Carey}, {Poleski}, {Skowron}, {Koz{\l}owski}, {Mr{\'o}z},
  {Pietrukowicz}, {Pietrzy{\'n}ski}, {Szyma{\'n}ski}, {Soszy{\'n}ski},
  {Ulaczyk}, {Wyrzykowski}, {OGLE Collaboration}, {Albrow}, {Beaulieu},
  {Caldwell}, {Cassan}, {Coutures}, {Danielski}, {Dominis Prester},
  {Donatowicz}, {Lon{\v c}ari{\'c}}, {McDougall}, {Morales}, {Ranc}, {Zhu},
  {PLANET Collaboration}, {Abe}, {Barry}, {Bennett}, {Bhattacharya},
  {Fukunaga}, {Inayama}, {Koshimoto}, {Namba}, {Sumi}, {Suzuki}, {Tristram},
  {Wakiyama}, {Yonehara}, {The MOA Collaboration}, {Maoz}, {Kaspi},
  {Friedmann}, {Wise Group}, {Bachelet}, {Figuera Jaimes}, {Bramich},
  {Tsapras}, {Horne}, {Snodgrass}, {Wambsganss}, {Steele}, {Kains}, {RoboNet
  Collaboration}, {Bozza}, {Dominik}, {J{\o}rgensen}, {Alsubai}, {Ciceri},
  {D'Ago}, {Haugb{\o}lle}, {Hessman}, {Hinse}, {Juncher}, {Korhonen},
  {Mancini}, {Popovas}, {Rabus}, {Rahvar}, {Scarpetta}, {Schmidt}, {Skottfelt},
  {Southworth}, {Starkey}, {Surdej}, {Wertz}, {Zarucki}, {MiNDSTEp Consortium},
  {Gaudi}, {Pogge}, {DePoy}, \& {{$\mu$}FUN Collaboration}}]{Calchi.2015.A}
{Calchi Novati}, S., {et~al.} 2015{\natexlab{a}}, \apj, 804, 20

\bibitem[{{Calchi Novati} {et~al.}(2015{\natexlab{b}}){Calchi Novati}, {Gould},
  {Yee}, {Beichman}, {Bryden}, {Carey}, {Fausnaugh}, {Gaudi}, {Henderson},
  {Pogge}, {Shvartzvald}, {Wibking}, {Zhu}, {Spitzer Team}, {Udalski},
  {Poleski}, {Pawlak}, {Szyma{\'n}ski}, {Skowron}, {Mr{\'o}z}, {Koz{\l}owski},
  {Wyrzykowski}, {Pietrukowicz}, {Pietrzy{\'n}ski}, {Soszy{\'n}ski}, {Ulaczyk},
  \& {OGLE Group}}]{Calchi.2015.B}
---. 2015{\natexlab{b}}, \apj, 814, 92

\bibitem[{{Claret}(2000)}]{Claret.2000.A}
{Claret}, A. 2000, \aap, 363, 1081

\bibitem[{{Dominik}(1999)}]{Dominik.1999.A}
{Dominik}, M. 1999, \aap, 349, 108

\bibitem[{{Dong} {et~al.}(2007){Dong}, {Udalski}, {Gould}, {Reach}, {Christie},
  {Boden}, {Bennett}, {Fazio}, {Griest}, {Szyma{\'n}ski}, {Kubiak},
  {Soszy{\'n}ski}, {Pietrzy{\'n}ski}, {Szewczyk}, {Wyrzykowski}, {Ulaczyk},
  {Wieckowski}, {Paczy{\'n}ski}, {DePoy}, {Pogge}, {Preston}, {Thompson}, \&
  {Patten}}]{Dong.2007.A}
{Dong}, S., {et~al.} 2007, \apj, 664, 862

\bibitem[{{Dong} {et~al.}(2009){Dong}, {Gould}, {Udalski}, {Anderson},
  {Christie}, {Gaudi}, {OGLE Collaboration}, {Jaroszy{\'n}ski}, {Kubiak},
  {Szyma{\'n}ski}, {Pietrzy{\'n}ski}, {Soszy{\'n}ski}, {Szewczyk}, {Ulaczyk},
  {Wyrzykowski}, {{$\mu$}FUN Collaboration}, {DePoy}, {Fox}, {Gal-Yam}, {Han},
  {L{\'e}pine}, {McCormick}, {Ofek}, {Park}, {Pogge}, {MOA Collaboration},
  {Abe}, {Bennett}, {Bond}, {Britton}, {Gilmore}, {Hearnshaw}, {Itow},
  {Kamiya}, {Kilmartin}, {Korpela}, {Masuda}, {Matsubara}, {Motomura},
  {Muraki}, {Nakamura}, {Ohnishi}, {Okada}, {Rattenbury}, {Saito}, {Sako},
  {Sasaki}, {Sullivan}, {Sumi}, {Tristram}, {Yanagisawa}, {Yock}, {Yoshoika},
  {PLANET/RoboNet Collaborations}, {Albrow}, {Beaulieu}, {Brillant}, {Calitz},
  {Cassan}, {Cook}, {Coutures}, {Dieters}, {Prester}, {Donatowicz},
  {Fouqu{\'e}}, {Greenhill}, {Hill}, {Hoffman}, {Horne}, {J{\o}rgensen},
  {Kane}, {Kubas}, {Marquette}, {Martin}, {Meintjes}, {Menzies}, {Pollard},
  {Sahu}, {Vinter}, {Wambsganss}, {Williams}, {Bode}, {Bramich}, {Burgdorf},
  {Snodgrass}, {Steele}, {Doublier}, \& {Foellmi}}]{Dong.2009.B}
---. 2009, \apj, 695, 970

\bibitem[{{Gould}(1994)}]{Gould.1994.A}
{Gould}, A. 1994, \apjl, 421, L75

\bibitem[{{Gould}(2008)}]{Gould.2008.A}
---. 2008, \apj, 681, 1593

\bibitem[{{Gould}(2013)}]{Gould.2013.A}
---. 2013, \apjl, 763, L35

\bibitem[{{Gould} \& {Gaucherel}(1997)}]{Gould.1997.A}
{Gould}, A., \& {Gaucherel}, C. 1997, \apj, 477, 580

\bibitem[{{Gould} \& {Horne}(2013)}]{Gould.2013.B}
{Gould}, A., \& {Horne}, K. 2013, \apjl, 779, L28

\bibitem[{{Gould} \& {Yee}(2012)}]{Gould.2012.A}
{Gould}, A., \& {Yee}, J.~C. 2012, \apjl, 755, L17

\bibitem[{{Gould} {et~al.}(2009){Gould}, {Udalski}, {Monard}, {Horne}, {Dong},
  {Miyake}, {Sahu}, {Bennett}, {Wyrzykowski}, {Soszy{\'n}ski}, {Szyma{\'n}ski},
  {Kubiak}, {Pietrzy{\'n}ski}, {Szewczyk}, {Ulaczyk}, {OGLE Collaboration},
  {Allen}, {Christie}, {DePoy}, {Gaudi}, {Han}, {Lee}, {McCormick}, {Natusch},
  {Park}, {Pogge}, {{$\mu$}FUN Collaboration}, {Allan}, {Bode}, {Bramich},
  {Burgdorf}, {Dominik}, {Fraser}, {Kerins}, {Mottram}, {Snodgrass}, {Steele},
  {Street}, {Tsapras}, {RoboNet Collaboration}, {Abe}, {Bond}, {Botzler},
  {Fukui}, {Furusawa}, {Hearnshaw}, {Itow}, {Kamiya}, {Kilmartin}, {Korpela},
  {Lin}, {Ling}, {Masuda}, {Matsubara}, {Muraki}, {Nagaya}, {Ohnishi},
  {Okumura}, {Perrott}, {Rattenbury}, {Saito}, {Sako}, {Skuljan}, {Sullivan},
  {Sumi}, {Sweatman}, {Tristram}, {Yock}, {MOA Collaboration}, {Albrow},
  {Beaulieu}, {Coutures}, {Calitz}, {Caldwell}, {Fouque}, {Martin}, {Williams},
  \& {PLANET Collaboration}}]{Gould.2009.B}
{Gould}, A., {et~al.} 2009, \apjl, 698, L147

\bibitem[{{Green} {et~al.}(2015){Green}, {Schlafly}, {Finkbeiner}, {Rix},
  {Martin}, {Burgett}, {Draper}, {Flewelling}, {Hodapp}, {Kaiser}, {Kudritzki},
  {Magnier}, {Metcalfe}, {Price}, {Tonry}, \& {Wainscoat}}]{Green.2015.A}
{Green}, G.~M., {et~al.} 2015, \apj, 810, 25

\bibitem[{{Grether} \& {Lineweaver}(2006)}]{Grether.2006.A}
{Grether}, D., \& {Lineweaver}, C.~H. 2006, \apj, 640, 1051

\bibitem[{{Han} \& {Gaudi}(2008)}]{Han.2008.A}
{Han}, C., \& {Gaudi}, B.~S. 2008, \apj, 689, 53

\bibitem[{{Han} {et~al.}(2016){Han}, {Jung}, {Udalski}, {Gould}, {Bozza},
  {Szyma{\'n}ski}, {Soszy{\'n}ski}, {Poleski}, {Koz{\l}owski}, {Pietrukowicz},
  {Skowron}, {Ulaczyk}, \& {Wyrzykowski}}]{Han.2016arXiv.A}
{Han}, C., {et~al.} 2016, ArXiv e-prints

\bibitem[{{Harps{\o}e} {et~al.}(2012){Harps{\o}e}, {J{\o}rgensen}, {Andersen},
  \& {Grundahl}}]{Harpsoe.2012.A}
{Harps{\o}e}, K.~B.~W., {J{\o}rgensen}, U.~G., {Andersen}, M.~I., \&
  {Grundahl}, F. 2012, \aap, 542, A23

\bibitem[{{Henderson} {et~al.}(2015){Henderson}, {Penny}, {Street}, {Bennett},
  {Hogg}, {Poleski}, {Barclay}, {Barentsen}, {Howell}, {Udalski},
  {Szyma{\'n}ski}, {Skowron}, {Mr{\'o}z}, {Koz{\l}owski}, {Wyrzykowski},
  {Pietrukowicz}, {Soszy{\'n}ski}, {Ulaczyk}, {Pawlak}, {Sumi}, {Abe},
  {Asakura}, {Barry}, {Bhattacharya}, {Bond}, {Donachie}, {Freeman}, {Fukui},
  {Hirao}, {Itow}, {Koshimoto}, {Li}, {Ling}, {Masuda}, {Matsubara}, {Muraki},
  {Nagakane}, {Ohnishi}, {Oyokawa}, {Rattenbury}, {Saito}, {Sharan},
  {Sullivan}, {Tristram}, {Yonehara}, {Bachelet}, {Bramich}, {Cassan},
  {Dominik}, {Figuera Jaimes}, {Horne}, {Hundertmark}, {Mao}, {Ranc},
  {Schmidt}, {Snodgrass}, {Steele}, {Tsapras}, {Wambsganss}, {Akeson},
  {Batista}, {Beaulieu}, {Beichman}, {Bozza}, {Bryden}, {Ciardi}, {Cole},
  {Coutures}, {Dong}, {Foreman-Mackey}, {Fouqu{\'e}}, {Gaudi}, {Kerins},
  {Korhonen}, {J{\o}rgensen}, {Lang}, {Lineweaver}, {Marquette}, {Mogavero},
  {Morales}, {Nataf}, {Pogge}, {Santerne}, {Shvartzvald}, {Suzuki}, {Tamura},
  {Tisserand}, {Wang}, \& {Zhu}}]{Henderson.2015arXiv.A}
{Henderson}, C.~B., {et~al.} 2015, ArXiv e-prints

\bibitem[{{Honma}(1999)}]{Honma.1999.A}
{Honma}, M. 1999, \apjl, 517, L35

\bibitem[{{Kervella} {et~al.}(2004){Kervella}, {Th{\'e}venin}, {Di Folco}, \&
  {S{\'e}gransan}}]{Kervella.2004.A}
{Kervella}, P., {Th{\'e}venin}, F., {Di Folco}, E., \& {S{\'e}gransan}, D.
  2004, \aap, 426, 297

\bibitem[{{Mogavero} \& {Beaulieu}(2016)}]{Mogavero.2016.A}
{Mogavero}, F., \& {Beaulieu}, J.~P. 2016, \aap, 585, A62

\bibitem[{{Muraki} {et~al.}(2011){Muraki}, {Han}, {Bennett}, {Suzuki},
  {Monard}, {Street}, {Jorgensen}, {Kundurthy}, {Skowron}, {Becker}, {Albrow},
  {Fouqu{\'e}}, {Heyrovsk{\'y}}, {Barry}, {Beaulieu}, {Wellnitz}, {Bond},
  {Sumi}, {Dong}, {Gaudi}, {Bramich}, {Dominik}, {Abe}, {Botzler}, {Freeman},
  {Fukui}, {Furusawa}, {Hayashi}, {Hearnshaw}, {Hosaka}, {Itow}, {Kamiya},
  {Korpela}, {Kilmartin}, {Lin}, {Ling}, {Makita}, {Masuda}, {Matsubara},
  {Miyake}, {Nishimoto}, {Ohnishi}, {Perrott}, {Rattenbury}, {Saito},
  {Skuljan}, {Sullivan}, {Sweatman}, {Tristram}, {Wada}, {Yock}, {MOA
  Collaboration}, {Christie}, {DePoy}, {Gorbikov}, {Gould}, {Kaspi}, {Lee},
  {Mallia}, {Maoz}, {McCormick}, {Moorhouse}, {Natusch}, {Park}, {Pogge},
  {Polishook}, {Shporer}, {Thornley}, {Yee}, {{$\mu$}FUN Collaboration},
  {Allan}, {Browne}, {Horne}, {Kains}, {Snodgrass}, {Steele}, {Tsapras},
  {RoboNet Collaboration}, {Batista}, {Bennett}, {Brillant}, {Caldwell},
  {Cassan}, {Cole}, {Corrales}, {Coutures}, {Dieters}, {Dominis Prester},
  {Donatowicz}, {Greenhill}, {Kubas}, {Marquette}, {Martin}, {Menzies}, {Sahu},
  {Waldman}, {Williams}, {Zub}, {PLANET Collaboration}, {Bourhrous},
  {Matsuoka}, {Nagayama}, {Oi}, {Randriamanakoto}, {IRSF Observers}, {Bozza},
  {Burgdorf}, {Calchi Novati}, {Dreizler}, {Finet}, {Glitrup}, {Harps{\o}e},
  {Hinse}, {Hundertmark}, {Liebig}, {Maier}, {Mancini}, {Mathiasen}, {Rahvar},
  {Ricci}, {Scarpetta}, {Skottfelt}, {Surdej}, {Southworth}, {Wambsganss},
  {Zimmer}, {MiNDSTEp Consortium}, {Udalski}, {Poleski}, {Wyrzykowski},
  {Ulaczyk}, {Szyma{\'n}ski}, {Kubiak}, {Pietrzy{\'n}ski}, {Soszy{\'n}ski}, \&
  {OGLE Collaboration}}]{Muraki.2011.A}
{Muraki}, Y., {et~al.} 2011, \apj, 741, 22

\bibitem[{{Nataf} {et~al.}(2013){Nataf}, {Gould}, {Fouqu{\'e}}, {Gonzalez},
  {Johnson}, {Skowron}, {Udalski}, {Szyma{\'n}ski}, {Kubiak},
  {Pietrzy{\'n}ski}, {Soszy{\'n}ski}, {Ulaczyk}, {Wyrzykowski}, \&
  {Poleski}}]{Nataf.2013.A}
{Nataf}, D.~M., {et~al.} 2013, \apj, 769, 88

\bibitem[{{Pejcha} \& {Heyrovsk{\'y}}(2009)}]{Pejcha.2009.A}
{Pejcha}, O., \& {Heyrovsk{\'y}}, D. 2009, \apj, 690, 1772

\bibitem[{{Ranc} {et~al.}(2015){Ranc}, {Cassan}, {Albrow}, {Kubas}, {Bond},
  {Batista}, {Beaulieu}, {Bennett}, {Dominik}, {Dong}, {Fouqu{\'e}}, {Gould},
  {Greenhill}, {J{\o}rgensen}, {Kains}, {Menzies}, {Sumi}, {Bachelet},
  {Coutures}, {Dieters}, {Dominis Prester}, {Donatowicz}, {Gaudi}, {Han},
  {Hundertmark}, {Horne}, {Kane}, {Lee}, {Marquette}, {Park}, {Pollard},
  {Sahu}, {Street}, {Tsapras}, {Wambsganss}, {Williams}, {Zub}, {Abe}, {Fukui},
  {Itow}, {Masuda}, {Matsubara}, {Muraki}, {Ohnishi}, {Rattenbury}, {Saito},
  {Sullivan}, {Sweatman}, {Tristram}, {Yock}, \& {Yonehara}}]{Ranc.2015.A}
{Ranc}, C., {et~al.} 2015, \aap, 580, A125

\bibitem[{{Refsdal}(1966)}]{Refsdal.1966.A}
{Refsdal}, S. 1966, \mnras, 134, 315

\bibitem[{{Schechter} {et~al.}(1993){Schechter}, {Mateo}, \&
  {Saha}}]{Schechter.1993.A}
{Schechter}, P.~L., {Mateo}, M., \& {Saha}, A. 1993, \pasp, 105, 1342

\bibitem[{{Shvartzvald} {et~al.}(2015){Shvartzvald}, {Udalski}, {Gould}, {Han},
  {Bozza}, {Friedmann}, {Hundertmark}, {and}, {Beichman}, {Bryden}, {Calchi
  Novati}, {Carey}, {Fausnaugh}, {Gaudi}, {Henderson}, {Kerr}, {Pogge},
  {Varricatt}, {Wibking}, {Yee}, {Zhu}, {Spitzer Team}, {Poleski}, {Pawlak},
  {Szyma{\'n}ski}, {Skowron}, {Mr{\'o}z}, {Koz{\l}owski}, {Wyrzykowski},
  {Pietrukowicz}, {Pietrzy{\'n}ski}, {Soszy{\'n}ski}, {Ulaczyk}, {OGLE Group},
  {Choi}, {Park}, {Jung}, {Shin}, {Albrow}, {Park}, {Kim}, {Lee}, {Cha}, {Kim},
  {Lee}, {KMTNet Group}, {Maoz}, {Kaspi}, {Wise Group}, {Street}, {Tsapras},
  {Bachelet}, {Dominik}, {Bramich}, {Horne}, {Snodgrass}, {Steele}, {Menzies},
  {Figuera Jaimes}, {Wambsganss}, {Schmidt}, {Cassan}, {Ranc}, {Mao}, {Dong},
  {RoboNet}, {D'Ago}, {Scarpetta}, {Verma}, {J{\o}rgensen}, {Kerins},
  {Skottfelt}, \& {MiNDSTEp}}]{Shvartzvald.2015.A}
{Shvartzvald}, Y., {et~al.} 2015, \apj, 814, 111

\bibitem[{{Shvartzvald} {et~al.}(2016){Shvartzvald}, {Maoz}, {Udalski}, {Sumi},
  {Friedmann}, {Kaspi}, {Poleski}, {Szyma{\'n}ski}, {Skowron}, {Koz{\l}owski},
  {Wyrzykowski}, {Mr{\'o}z}, {Pietrukowicz}, {Pietrzy{\'n}ski},
  {Soszy{\'n}ski}, {Ulaczyk}, {Abe}, {Barry}, {Bennett}, {Bhattacharya},
  {Bond}, {Freeman}, {Inayama}, {Itow}, {Koshimoto}, {Ling}, {Masuda}, {Fukui},
  {Matsubara}, {Muraki}, {Ohnishi}, {Rattenbury}, {Saito}, {Sullivan},
  {Suzuki}, {Tristram}, {Wakiyama}, \& {Yonehara}}]{Shvartzvald.2016.A}
---. 2016, \mnras, 457, 4089

\bibitem[{{Skottfelt} {et~al.}(2015){Skottfelt}, {Bramich}, {Hundertmark},
  {J{\o}rgensen}, {Michaelsen}, {Kj{\ae}rgaard}, {Southworth}, {S{\o}rensen},
  {Andersen}, {Andersen}, {Christensen-Dalsgaard}, {Frandsen}, {Grundahl},
  {Harps{\o}e}, {Kjeldsen}, \& {Pall{\'e}}}]{Skottfelt.2015.A}
{Skottfelt}, J., {et~al.} 2015, \aap, 574, A54

\bibitem[{{Street} {et~al.}(2016){Street}, {Udalski}, {Calchi Novati},
  {Hundertmark}, {Zhu}, {Gould}, {Yee}, {Tsapras}, {Bennett}, {RoboNet
  Project}, {Consortium}, {J{\o}rgensen}, {Dominik}, {Andersen}, {Bachelet},
  {Bozza}, {Bramich}, {Burgdorf}, {Cassan}, {Ciceri}, {D'Ago}, {Dong}, {Evans},
  {Gu}, {Harkonnen}, {Hinse}, {Horne}, {Figuera Jaimes}, {Kains}, {Kerins},
  {Korhonen}, {Kuffmeier}, {Mancini}, {Menzies}, {Mao}, {Peixinho}, {Popovas},
  {Rabus}, {Rahvar}, {Ranc}, {Tronsgaard Rasmussen}, {Scarpetta}, {Schmidt},
  {Skottfelt}, {Snodgrass}, {Southworth}, {Steele}, {Surdej}, {Unda-Sanzana},
  {Verma}, {von Essen}, {Wambsganss}, {Wang}, {Wertz}, {OGLE Project},
  {Poleski}, {Pawlak}, {Szyma{\'n}ski}, {Skowron}, {Mr{\'o}z}, {Koz{\l}owski},
  {Wyrzykowski}, {Pietrukowicz}, {Pietrzy{\'n}ski}, {Soszy{\'n}ski}, {Ulaczyk},
  {Spitzer Team}, {Beichman}, {Bryden}, {Carey}, {Gaudi}, {Henderson}, {Pogge},
  {Shvartzvald}, {The MOA Collaboration}, {Abe}, {Asakura}, {Bhattacharya},
  {Bond}, {Donachie}, {Freeman}, {Fukui}, {Hirao}, {Inayama}, {Itow},
  {Koshimoto}, {Li}, {Ling}, {Masuda}, {Matsubara}, {Muraki}, {Nagakane},
  {Nishioka}, {Ohnishi}, {Oyokawa}, {Rattenbury}, {Saito}, {Sharan},
  {Sullivan}, {Sumi}, {Suzuki}, {Tristram}, {Wakiyama}, {Yonehara}, {KMTNet
  Modeling Team}, {Han}, {Choi}, {Park}, {Jung}, \& {Shin}}]{Street.2016.A}
{Street}, R.~A., {et~al.} 2016, \apj, 819, 93

\bibitem[{{Sumi} {et~al.}(2003){Sumi}, {Abe}, {Bond}, {Dodd}, {Hearnshaw},
  {Honda}, {Honma}, {Kan-ya}, {Kilmartin}, {Masuda}, {Matsubara}, {Muraki},
  {Nakamura}, {Nishi}, {Noda}, {Ohnishi}, {Petterson}, {Rattenbury}, {Reid},
  {Saito}, {Saito}, {Sato}, {Sekiguchi}, {Skuljan}, {Sullivan}, {Takeuti},
  {Tristram}, {Wilkinson}, {Yanagisawa}, \& {Yock}}]{Sumi.2003.A}
{Sumi}, T., {et~al.} 2003, \apj, 591, 204

\bibitem[{{Udalski}(2003)}]{Udalski.2003.A}
{Udalski}, A. 2003, \actaa, 53, 291

\bibitem[{{Udalski} {et~al.}(2015{\natexlab{a}}){Udalski}, {Szyma{\'n}ski}, \&
  {Szyma{\'n}ski}}]{Udalski.2015.B}
{Udalski}, A., {Szyma{\'n}ski}, M.~K., \& {Szyma{\'n}ski}, G.
  2015{\natexlab{a}}, \actaa, 65, 1

\bibitem[{{Udalski} {et~al.}(2015{\natexlab{b}}){Udalski}, {Yee}, {Gould},
  {Carey}, {Zhu}, {Skowron}, {Koz{\l}owski}, {Poleski}, {Pietrukowicz},
  {Pietrzy{\'n}ski}, {Szyma{\'n}ski}, {Mr{\'o}z}, {Soszy{\'n}ski}, {Ulaczyk},
  {Wyrzykowski}, {Han}, {Calchi Novati}, \& {Pogge}}]{Udalski.2015.A}
{Udalski}, A., {et~al.} 2015{\natexlab{b}}, \apj, 799, 237

\bibitem[{{Yee} {et~al.}(2016){Yee}, {Johnson}, {Skowron}, {Gould}, {Pineda},
  {Eastman}, {Vanderburg}, \& {Howard}}]{Yee.2016.A}
{Yee}, J.~C., {Johnson}, J.~A., {Skowron}, J., {Gould}, A., {Pineda}, J.~S.,
  {Eastman}, J., {Vanderburg}, A., \& {Howard}, A. 2016, \apj, 821, 121

\bibitem[{{Yee} {et~al.}(2015){Yee}, {Gould}, {Beichman}, {Calchi Novati},
  {Carey}, {Gaudi}, {Henderson}, {Nataf}, {Penny}, {Shvartzvald}, \&
  {Zhu}}]{Yee.2015.A}
{Yee}, J.~C., {et~al.} 2015, \apj, 810, 155

\bibitem[{{Zhu} {et~al.}(2015{\natexlab{a}}){Zhu}, {Calchi Novati}, {Gould},
  {Udalski}, {Han}, {Shvartzvald}, {Ranc}, {Jorgensen}, {Poleski}, {Bozza},
  {Beichman}, {Bryden}, {Carey}, {Gaudi}, {Henderson}, {Pogge}, {Porritt},
  {Wibking}, {Yee}, {Pawlak}, {Szymanski}, {Skowron}, {Mroz}, {Kozlowski},
  {Wyrzykowski}, {Pietrukowicz}, {Pietrzynski}, {Soszynski}, {Ulaczyk}, {Choi},
  {Park}, {Jung}, {Shin}, {Albrow}, {Park}, {Kim}, {Lee}, {Kim}, {Lee},
  {Friedmann}, {Kaspi}, {Maoz}, {Hundertmark}, {Street}, {Tsapras}, {Bramich},
  {Cassan}, {Dominik}, {Bachelet}, {Dong}, {Figuera Jaimes}, {Horne}, {Mao},
  {Menzies}, {Schmidt}, {Snodgrass}, {Steele}, {Wambsganss}, {Skottfelt},
  {Andersen}, {Burgdorf}, {Ciceri}, {D'Ago}, {Evans}, {Gu}, {Hinse}, {Kerins},
  {Korhonen}, {Kuffmeier}, {Mancini}, {Peixinho}, {popovas}, {Rabus}, {Rahvar},
  {Rasmussen}, {Scarpetta}, {Southworth}, {Surdej}, {von Essen}, {Wang}, \&
  {Wertz}}]{Zhu.2015arXiv.A}
{Zhu}, W., {et~al.} 2015{\natexlab{a}}, ArXiv e-prints

\bibitem[{{Zhu} {et~al.}(2015{\natexlab{b}}){Zhu}, {Gould}, {Beichman}, {Calchi
  Novati}, {Carey}, {Gaudi}, {Henderson}, {Penny}, {Shvartzvald}, {Yee},
  {Udalski}, {Poleski}, {Skowron}, {Koz{\l}owski}, {Mr{\'o}z}, {Pietrukowicz},
  {Pietrzy{\'n}ski}, {Szyma{\'n}ski}, {Soszy{\'n}ski}, {Ulaczyk},
  {Wyrzykowski}, {OGLE Collaboration}, {Abe}, {Barry}, {Bennett},
  {Bhattacharya}, {Bond}, {Freeman}, {Fukui}, {Hirao}, {Itow}, {Koshimoto},
  {Ling}, {Masuda}, {Matsubara}, {Muraki}, {Nagakane}, {Ohnishi}, {Saito},
  {Sullivan}, {Sumi}, {Suzuki}, {Tristram}, {Rattenbury}, {Wakiyama},
  {Yonehara}, {The MOA Collaboration}, {Maoz}, {Kaspi}, {Friedmann}, \& {The
  Wise Group}}]{Zhu.2015.A}
---. 2015{\natexlab{b}}, \apj, 814, 129

\bibitem[{{Zhu} {et~al.}(2015{\natexlab{c}}){Zhu}, {Udalski}, {Gould},
  {Dominik}, {Bozza}, {Han}, {Yee}, {Calchi Novati}, {Beichman}, {Carey},
  {Poleski}, {Skowron}, {Koz{\l}owski}, {Mr{\'o}z}, {Pietrukowicz},
  {Pietrzy{\'n}ski}, {Szyma{\'n}ski}, {Soszy{\'n}ski}, {Ulaczyk},
  {Wyrzykowski}, {OGLE Collaboration}, {Gaudi}, {Pogge}, {DePoy}, {Jung},
  {Choi}, {Hwang}, {Shin}, {Park}, {Jeong}, \& {{$\mu$}FUN
  Collaboration}}]{Zhu.2015.B}
---. 2015{\natexlab{c}}, \apj, 805, 8

\end{thebibliography}
\end{document}